\documentclass[10pt,showpacs,prd,preprintnumbers,nofootinbib,superscriptaddress,nobibnotes,
]{revtex4-2}
\usepackage{subcaption}
\usepackage[export]{adjustbox}
\usepackage[font=normalsize,labelfont=bf]{caption}

\usepackage{caption}
\usepackage{amsmath,amsthm,amssymb}
\usepackage{multirow}
\usepackage{mathrsfs} 
\usepackage{tensor} 
\usepackage[normalem]{ulem} 
\usepackage{graphicx}
\usepackage{latexsym}
\usepackage{url,hyperref}
\usepackage{textcomp}
\usepackage[dvipsnames]{xcolor} 
\usepackage{slashed}
\usepackage{placeins}
\usepackage{array}
\usepackage{cancel}
\usepackage{verbatim}
\usepackage{caption}
\usepackage{subcaption}
\usepackage[splitrule,hang,flushmargin]{footmisc} 
\renewcommand{\footnoterule}{%
  \kern -3pt
  \hrule width \textwidth height 1pt
  \kern 2pt
}

\usepackage{chngcntr}
\usepackage{pdfpages}

\usepackage{booktabs}
\usepackage{geometry}
\usepackage{multirow}
\usepackage{setspace} 
\AtBeginDocument{\setstretch{1.125}}

\usepackage{outlines} 
\usepackage{csquotes} 

\usepackage{blindtext}
\usepackage{tcolorbox}

\hypersetup{colorlinks=true,linkcolor=blue,citecolor=magenta,filecolor=magenta,urlcolor=blue}

\numberwithin{equation}{section}

\DeclareUnicodeCharacter{2212}{-} 

\usepackage{etoolbox}

\makeatletter
\patchcmd{\@outputpage@head}{\@ifx{\LS@rot\@undefined}{}{\LS@rot}}{}{}{}
\patchcmd{\frontmatter@abstract@produce} 
  {\vskip200\p@\@plus1fil
   \penalty-200\relax
   \vskip-200\p@\@plus-1fil}
  {}
  {}
  {}
\makeatother

\newcolumntype{C}{>{$}c<{$}} 
\newcolumntype{L}{>{$}l<{$}} 

\renewcommand{\arraystretch}{1.5} 

\begin{document}

\title{Gravitationally-decoupled hairy black holes:
probing geometric, optical, and quasinormal mode signatures}

\author{Paul Allen M. Gonzales}
\email{paul\_allen\_gonzales@dlsu.edu.ph}
\affiliation{Department of Physics, De La Salle University, 2401 Taft Avenue, Manila, 1004 Philippines}

\author{Anna Chrysostomou}
\email{chrysostomou@lpthe.jussieu.fr}
\affiliation{Laboratoire de Physique Th\'eorique et Hautes \'Energies - LPTHE, Sorbonne Universit\'e, CNRS, 4 Place Jussieu, 75005 Paris, France}

\author{Alan S. Cornell}
\email{acornell@uj.ac.za}
\affiliation{Department of Physics, University of Johannesburg, PO Box 524, Auckland Park 2006, South Africa.}
\affiliation{Department of Physics, De La Salle University, 2401 Taft Avenue, Manila, 1004 Philippines}

\author{Emmanuel Rodulfo}
\email{emmanuel.rodulfo@dlsu.edu.ph}
\affiliation{Department of Physics, De La Salle University, 2401 Taft Avenue, Manila, 1004 Philippines}

\begin{abstract}
Gravitational-decoupling provides a systematic framework for constructing black hole geometries from known (seed) solutions (such as the Schwarzschild), whose exterior properties closely resemble those of the seed solution while still retaining potentially distinguishable features. Using this approach, we study a class of static, spherically symmetric hairy black holes and examine how such deformations affect their horizon structure, photon spheres, shadow scales, and scalar quasinormal modes (QNMs). We first determine the physically controlled regions of the parameter space by imposing the relevant horizon and energy condition requirements, while retaining configurations outside these domains only for exploratory comparison. To isolate geometric effects, we compare solutions at a common horizon radius, while accounting explicitly for the distinction between the mass parameter entering the seed geometry and the ADM mass measured asymptotically. The resulting ADM normalized shadow scales are used for an illustrative comparison with the characteristic angular scales of M87* and Sgr~A*. We also compute massless scalar QNMs using the sixth-order Wentzel-Kramers-Brillouin approximation and the improved Asymptotic Iteration Method, finding close agreement over the modes considered. Our results show that fixing the horizon radius does not uniquely determine either the optical or perturbative properties of the spacetime, and that mass normalization can significantly alter the interpretation of geometric trends. Horizon structure, null geodesic observables, and scalar QNMs therefore provide complementary probes of gravitationally-decoupled black hole geometries.

\end{abstract}

\date{\today}
\maketitle

{
\begin{spacing}{1}
  \hypersetup{linkcolor=RoyalPurple}
\tableofcontents
\end{spacing}
}


\section{Introduction}\label{intro}
\par Black holes provide a laboratory in which to conduct some of the most stringent theoretical and observational tests of General Relativity (GR). The Schwarzschild solution \cite{Schwarzschild:1916uq}, its charged Reissner-Nordstr\"om (RN) generalization \cite{Reissner:1916cle,Nordstrom:2018acn}, and the rotating Kerr solution \cite{Kerr:1963ud} form the canonical families of asymptotically flat black hole spacetimes. Under the assumptions entering the standard black hole uniqueness theorems, stationary black holes in four-dimensional Einstein-Maxwell theory are characterized by their mass, angular momentum, and electromagnetic charge \cite{Israel:1967wq,Robinson:1975bv,Bekenstein:1995un}. The broader no-hair paradigm expresses the corresponding expectation that additional independent degrees of freedom should not generically survive outside the event horizon. This expectation depends, however, on the matter content, gravitational theory, and assumptions imposed on the solution. Numerous examples of black hole solutions with additional structure are now known. These include non-Abelian \enquote{colored} black holes in Einstein-Yang-Mills theory \cite{Bizon:1990sr,Volkov:1998cc}, scalar hair in scalar-tensor and Horndeski theories \cite{Sotiriou:2011dz,Sotiriou:2014pfa,Babichev:2013cya,Cisterna:2014nua}, scalarized solutions generated through non-minimal curvature couplings \cite{Antoniou:2017acq,Antoniou:2017hxj}, and rotating black holes supporting synchronized scalar hair \cite{Herdeiro2015}. This non-exhaustive set illustrates the variety of mechanisms through which additional fields or effective source sectors can modify a black hole exterior while preserving a regular event horizon.

\par Despite this broad theoretical landscape, establishing whether such additional structure can be distinguished observationally remains challenging. In Ref. \cite{zhou2017iron}, studies of X-ray reflection spectroscopy as a means of distinguishing Kerr black holes from Kerr black holes with Proca hair, found substantial degeneracy at the observational precision considered. Conversely, other studies have identified potentially distinguishable signatures in suitable regions of parameter space. For example, characteristic photon ring features have been studied for hairy black holes in an Einstein-Maxwell-scalar model \cite{Gan:2021xdl}, while a series of works explored how additional hair modifies the optical properties of black hole geometries \cite{Meng:2023htc,Meng:2024puu,Meng:2025ivb}. These results illustrate that the observational visibility of black hole hair is strongly dependent on both the underlying model and the observable under consideration.

\par The development of gravitational-wave astronomy \cite{LIGOScientific:2016aoc,LIGOScientific:2018mvr,LIGOScientific:2020ibl,KAGRA:2021vkt} and horizon-scale imaging by the Event Horizon Telescope (EHT) has opened complementary avenues for probing deviations from standard black hole geometries. The EHT observations of M87* and Sgr~A* provide access to the strong-field region surrounding supermassive black holes \cite{EventHorizonTelescope:2019dse,EventHorizonTelescope:2022wkp}. For static, spherically symmetric geometries, the unstable circular null orbit determines the photon sphere radius, while the associated critical impact parameter sets the characteristic geometric scale of the black hole shadow. This scale can be translated into a critical angular diameter once the physical mass and distance are specified. The geometrically defined shadow scale should, however, be distinguished from the observed emission ring diameter, whose detailed interpretation additionally depends on the emitting plasma and image modeling.

\par The quasinormal modes (QNMs) that dominate the post-merger phase of a black hole merger event provide a complementary probe of their black hole source \cite{Vishveshwara:1970zz,Press:1971wr,Chandrasekhar:1975zza,Detweiler:1977gy}. Their complex quasinormal frequencies (QNFs) encode characteristic oscillation and damping scales, and depend sensitively on the effective potential governing the perturbation. In the eikonal regime, the QNM spectrum is closely related to the properties of unstable null geodesics \cite{refGoebel1972}, establishing a useful connection between photon sphere structure and perturbative behavior. More generally, QNM properties can also be studied using scalar test-fields propagating on fixed black hole backgrounds, where such calculations provide a controlled probe of qualitative features of the perturbative response; in this way, the QNM spectrum of a scalar test-field serves as a proxy for (while nevertheless remaining distinct from) the full gravitational perturbation spectrum. This distinction is particularly important for hairy geometries, for which a complete stability analysis may require simultaneous perturbations of the metric and the additional source fields. Photon sphere and shadow observables therefore probe aspects of the spacetime that are different from those probed by QNM spectra; employing a variety of probes may help distinguish geometries that remain degenerate under any single diagnostic.

\par This motivates a complementary approach in which controlled deformations of a reference black hole geometry are studied directly through quantities accessible outside the horizon. Effective and parametrized descriptions of black hole geometries have been developed for this purpose. For example, the Effective Metric Description \cite{Binetti:2022xdi,DamiaPaciarini:2025xjc} parametrizes deformations of static, spherically symmetric Schwarzschild geometries in terms of physical quantities defined relative to the horizon, and has subsequently been extended to observables including the black hole shadow \cite{DelPiano:2024nrl}. Related recent work \cite{Ylla:2026ffv} investigated perturbations of hairy Schwarzschild and Kerr black holes by modeling the hair through an anisotropic-fluid deformation and exploiting the correspondence between QNMs and null geodesics \cite{refGoebel1972,refCardosoLyapunov}. 

\par In the present work, we investigate these questions within the gravitational-decoupling framework \cite{Ovalle:2016pwp,Ovalle:2017fgl,Ovalle:2018umz,Ovalle:2018gic,Ovalle:2020kpd}. Gravitational-decoupling provides a systematic means of generating static, spherically symmetric solutions by supplementing a known seed stress-energy tensor with an additional source sector $\Theta_{\mu\nu}$, and correspondingly deforming the seed geometry. Depending on the construction, physical restrictions may then be imposed through the horizon structure and appropriate energy conditions. The resulting analytic geometries therefore provide useful laboratories in which the influence of an additional source sector can be followed explicitly from the metric to its null geodesic and perturbative properties. As such, we shall focus on four analytic subfamilies of the gravitationally-decoupled solutions introduced by Ovalle \textit{et al.}, denoted as \enquote{hairy metrics} HM1--HM4. Since the phenomenological utility of these geometries depends first on establishing that the corresponding configurations are physically meaningful, a central part of our analysis is the identification of controlled regions of parameter space. We impose the relevant horizon and energy condition requirements, and distinguish physically admissible configurations from those retained only for exploratory comparison. The four geometries differ in the status of their admissible domains, reflecting their distinct constructions under specific energy conditions. Under the criteria adopted here, HM1--HM3 possess controlled parameter ranges; upon failing to meet a specific energy condition at the horizon, HM4 configurations are retained solely as exploratory geometric and perturbative benchmarks.

\par To compare the different geometries on a common footing, we employ a fixed-horizon criterion that sets the event horizon as $r_{\rm h}=2M$, where appropriate. This provides a common benchmark in which the horizon radius is fixed relative to the seed mass $M$. We discuss how this choice does not correspond to a comparison at fixed physical mass, since the Arnowitt-Deser-Misner (ADM) mass generally varies with the deformation parameters. We therefore retain the seed-mass normalization for the geometric and QNM calculations, while explicitly accounting for the ADM mass when translating the critical impact parameter into an astrophysical angular scale. Following the geometric shadow and lensing-motivated analyses of Refs.~\cite{Crisnejo:2019ril,mustafa2024testing}, we then compare the resulting critical angular diameters with the characteristic EHT angular scales of M87* and Sgr~A*. These comparisons are intended to illustrate the sensitivity of the geometric shadow scale to the different deformations and are not interpreted as direct statistical constraints on the associated hair parameters.

\par For the perturbative analysis, we consider a massless, minimally coupled scalar test-field on each fixed black hole background. The QNFs are calculated using two semi-analytic techniques: the sixth-order Wentzel-Kramers-Brillouin (WKB) approximation \cite{Konoplya:2004ip} and the improved Asymptotic Iteration Method (AIM) \cite{Cho:2009cj,Cho:2011sf}. In addition to applying the WKB method to the effective potentials of the four geometries, we derive a common improved AIM formulation in a compact radial coordinate after explicitly factoring the QNM boundary behavior. We compute the QNFs using both methods, under the fixed-horizon criterion. We also consider additional exploratory QNM configurations beyond the $r_{\rm h}=2M$ benchmark in order to examine whether fixing the horizon radius is sufficient to characterize the perturbative response.

\par Our aim in this work is therefore twofold: first, to establish which regions of these gravitationally-decoupled black hole families admit a controlled physical interpretation under the horizon and energy condition criteria adopted here; second, to determine how the corresponding geometric deformations are reflected across a complementary set of exterior probes. We examine the horizon structure, photon sphere radii, critical impact parameters, ADM-normalized angular scales, effective scalar potentials, and scalar QNM spectra. This combined analysis allows us to assess which features arise generally from geometric deformation and which depend sensitively on the particular hairy black hole family. As such, this paper is organized as follows: Sec.~\ref{GD} reviews the gravitational-decoupling framework and introduces the four hairy black hole geometries. In Sec.~\ref{analysis of metric models}, we establish the corresponding parameter restrictions and analyze their horizon structure, photon sphere properties, shadow scales, and critical angular diameters. Sec.~\ref{QNMS} presents the scalar test-field QNM analysis using the sixth-order WKB method and the improved AIM. The combined geometric, optical, and perturbative results are discussed in Sec.~\ref{Discussion}, and our conclusions are summarized in Sec.~\ref{Conclusion}. 


\section{Gravitational decoupling framework and hairy black hole solutions}\label{GD}
\par To contextualize  our analysis and make our conventions explicit, we briefly review the construction of hairy black hole solutions within the gravitational decoupling framework (for a detailed discussion, see Ref.~\cite{Ovalle:2018gic}). Throughout, we work in four spacetime dimensions and adopt the metric signature $(-,+,+,+)$. We begin with Einstein's field equations,
\begin{equation}\label{eq:EFE}
G_{\mu\nu}=k^2\widetilde T_{\mu\nu} \;,
\qquad
k^2=8\pi G_N \;,
\end{equation}
where $G_{\mu\nu} = R_{\mu\nu} - Rg_{\mu\nu}/2$ is Einstein's tensor describing the geometry of the spacetime, $k$ is defined through $k^2 = 8\pi G_N$ with Newton's constant $G_N$, and $\widetilde{T}_{\mu\nu}$ is the total stress-energy tensor encoding the matter and energy of the spacetime. We work in natural units, such that $G_N=c=1$. 

\par In the gravitational decoupling framework,\footnote{Minimal Gravitational Decoupling \cite{Ovalle:2017fgl} is based on a deformation of the radial component, while Extended Gravitational Decoupling \cite{Ovalle:2018gic} is based on deformations of both radial and temporal components. Here, we adopt the latter because it provides a more general framework that encompasses both cases.} static and spherically symmetric black hole solutions are generated by decomposing the total stress-energy tensor into a seed source, $T_{\mu \nu}$, and an additional source, $\Theta_{\mu \nu}$ \cite{Ovalle:2016pwp,Ovalle:2017fgl,Ovalle:2018umz,Ovalle:2018gic,Ovalle:2020kpd}, such that
\begin{equation}\label{eq:TotalTmunu}
\widetilde{T}_{\mu\nu}=T_{\mu\nu}+\Theta_{\mu\nu} \;.
\end{equation}
The source $T_{\mu\nu}$ determines a known seed geometry, while $\Theta_{\mu\nu}$ represents an additional sector that may encode new matter fields, anisotropic stresses, or other effective gravitational degrees of freedom.

\par In this work, we adopt the static, spherically symmetric areal-radius gauge, with $g_{tt}g_{rr}=-1$, so that the solution to Eq.~\eqref{eq:EFE} corresponding to the total
stress-energy tensor
$\widetilde{T}_{\mu \nu}$ can be written as
\begin{equation}\label{eq:etakappametric}
ds_{\rm total}^2=-e^{\eta(r)}dt^2+ e^{\kappa(r)}dr^2 +r^2d\Omega^2 \;,
\end{equation}
where $d\Omega^2 = d\vartheta^2 + \sin^2\vartheta d \phi^2$, and the terms $\eta = \eta (r)$ and $\kappa = \kappa (r)$ depend only on the areal radius, $r$. The condition $e^{\eta}=e^{-\kappa}$ implies that the temporal and radial metric components are governed by a common metric function, whose zeros identify the corresponding Killing horizons.

\par Substituting Eq.~\eqref{eq:etakappametric} into Eq.~\eqref{eq:EFE} yields a coupled system of differential equations relating the spacetime geometry to the components of the total stress-energy tensor \cite{Ovalle:2020kpd},
\begin{equation}
    \widetilde T^{\mu}_{\nu} = {\rm diag} \left (-\widetilde \rho, \widetilde p_r, \widetilde p_t, \widetilde p_t \right), \qquad
    \widetilde{\rho} = -\left(T^0{}_0+\Theta^0{}_0\right) \;,
   \quad
   \widetilde{p}_r = T^1{}_1+
   \Theta^1{}_1 \;,
   \quad
   \widetilde{p}_t = T^2{}_2+\Theta^2{}_2  \;.
\end{equation}
From this system of equations, a non-vanishing anisotropy emerges,
\begin{equation}
   \Pi \equiv \widetilde{p}_t - \widetilde{p}_r \neq 0 \;,
\end{equation}
which allows for an anisotropic fluid treatment \cite{herrera1997local,mak2003anisotropic}.

\par The solution to Eq.~(\ref{eq:EFE}) associated only to the seed source $T_{\mu\nu}$ (i.e. when $\Theta_{\mu \nu}=0$) is
\begin{equation}
\label{eq:ximumetric}
ds_{\rm seed}^2 = -e^{\xi(r)} dt^2 + e^{\mu(r)} dr^2 + r^2 d\Omega^2 \;.
\end{equation}
Here, $e^{-\mu(r)}$ may be expressed in terms of the Misner-Sharp mass function $m=m(r)$ \cite{Misner:1964je}. 

\par To convey the influence of $\Theta_{\mu \nu}$ on the seed geometry, we introduce the extended metric deformations,
\begin{align}
    \xi \mapsto \eta &= \xi + \alpha \, h_1 \label{xi}\;,\\
    e^{-\mu} \mapsto e^{-\kappa} &= e^{-\mu} + \alpha \, h_2 \label{mu}\;,
\end{align}
where $h_1$ and $h_2$ describe the temporal and radial deformations, respectively. The dimensionless parameter $\alpha$ controls the strength of the geometric deformation induced by the additional source sector.

\par The Einstein equations corresponding to the total stress-energy tensor defined in Eq.~\eqref{eq:TotalTmunu} are then separated into two systems of equations: the first corresponds to the standard Einstein equations sourced by $T_{\mu\nu}$, while the second describes the additional sector $\Theta_{\mu\nu}$ in terms of the deformation functions introduced in Eqs.~\eqref{xi} and \eqref{mu}. This decomposition makes explicit how the deformation sector couples to the seed geometry and demonstrates that the limit $\alpha\rightarrow0$ smoothly recovers the original seed spacetime. This procedure follows the conservation equation,
\begin{equation}
\nabla_{\mu} \widetilde{T}^{\mu \nu} = 0 \;,
\end{equation}
corresponding to the metrics Eqs.~\eqref{eq:etakappametric} and \eqref{eq:ximumetric}; Ovalle \textit{et al.} demonstrate that $T_{\mu \nu}$ and $\Theta_{\mu \nu}$ can be decoupled without introducing a perturbative expansion in the parameter $\alpha$. Moreover, the framework of Ovalle \textit{et al.} leads to specific cases of hairy black hole solutions by imposing conditions on the additional source \textit{$\Theta_{\mu\nu}$}, such as the existence of well-defined event horizon and the satisfaction of the energy conditions.\footnote{For the effective anisotropic stress-energy tensor $\widetilde T^\mu{}_\nu$, the Strong and Dominant Energy Conditions take the form
\begin{equation*}
   \widetilde{\rho} + \widetilde{p_r} + 2\widetilde{p_t} \geq 0 \; , \quad
   \widetilde{\rho} + \widetilde{p_r} \geq 0 \; , \quad
   \widetilde{\rho} + \widetilde{p_t} \geq 0 \; ; \qquad
   \widetilde{\rho} \geq 0 \;, 
   \quad
   \widetilde{\rho} \geq |\widetilde{p_r}| \; , \quad
   \widetilde{\rho} \geq |\widetilde{p_t}| \; .
\end{equation*}
These conditions are imposed to ensure that the effective matter sector sourcing the spacetime remains physically reasonable and does not exhibit manifestly pathological energy densities or fluxes.} These requirements generate distinct analytic branches of solutions and introduce additional integration constants, denoted here by $j$ and, where relevant, an effective charge parameter $Q$. While the parameter $\alpha$ controls the strength of the geometric deformation sourced by $\Theta_{\mu\nu}$, $j$ parametrizes the corresponding hair sector. In several of the solutions considered below, the combination $\alpha j$ is the quantity that directly enters the horizon structure. The parameter $Q$ denotes an effective gauge charge associated with the same deformation sector.\footnote{The charge $Q$ need not correspond to an electromagnetic charge. Depending on the underlying source sector, it may instead encode the effects of additional matter fields, anisotropic stresses, higher-dimensional corrections, or other effective geometric contributions.} Throughout this work, we collectively refer to $\alpha$, $j$, and (where applicable) $Q$ as hair or deformation parameters. It should be emphasized that $\alpha$, $j$, and $Q$ are not generally independent parameters. In the analytic subfamilies considered below, the imposed energy conditions and horizon constraints generate nontrivial relations among these quantities and, in some cases, relate them directly to the seed mass parameter $M$.

\par We consider four specific subfamilies of these solutions, which we label as \enquote{Hairy Metric $i$} (HM$i$), with $i=1,2,3,4$. The first solution, HM1, arises from imposing the Strong Energy Condition (SEC) on a Schwarzschild seed geometry. The corresponding SEC branch yields a parent metric function of the form
\begin{equation}
f_{\rm SEC}(r) = 1-\frac{2\mathcal M}{r} + \alpha e^{-r/(\mathcal M-\alpha j/2)}  \; , \qquad \mathcal M = M+\frac{\alpha j}{2} \; .
\end{equation}
Imposing the additional requirement that the event horizon remain fixed at $r_{\rm h}=2M$ yields the first hairy black hole metric considered in this work,
\begin{equation}\label{eq:HM1}
\text{HM1:} \qquad f_{\rm HM1}(r)=1-\frac{2M}{r} +\alpha\left(e^{-r/M}-\frac{2M}{e^2r}\right) \; .
\end{equation}
This geometry is characterized by the seed mass $M$ and the deformation strength $\alpha$. The exponentially decaying contribution reflects the localization of the additional source near the black hole, ensuring that its influence becomes increasingly suppressed at large distances. The remaining three solutions arise from imposing the Dominant Energy Condition (DEC) on the corresponding decoupled system. Under this constraint, the generic family contains the parameters $\{M,\alpha,j,Q\}$. The parent metric function may be written as
\begin{equation} \label{eq:fDEC}
f_{\rm DEC}(r) = 1-\frac{2M+\alpha j}{r} + \frac{Q^2}{r^2} - \frac{\alpha M}{r}e^{-r/M} \; .
\end{equation}
Three analytic subclasses emerge upon imposing specific relations among $M$, $\alpha$, $j$, and $Q$.

\par For
\begin{equation}
Q^2 = 4\alpha\left(\frac{M}{e}\right)^2  \; , \qquad j = \frac{M}{e^2} \; ,
\end{equation}
we obtain
\begin{equation}\label{eq:fHM2}
\text{HM2:}\qquad f_{\rm HM2}(r) = 1-\frac{2M}{r}\left(1+\frac{\alpha}{2e^2}\right) + \frac{4\alpha M^2}{e^2r^2} -\frac{\alpha M}{r}e^{-r/M} \; ,
\end{equation}
with $r_{\rm h}=2M$.
In this subclass, $Q$, $M$, and $\alpha$ are not independent quantities. Consequently, varying $\alpha$ while holding $M$ fixed necessarily modifies the effective charge parameter $Q$.

\par For
\begin{equation}
Q^2=\alpha jM\left(2+\alpha e^{-\alpha j/M}\right) \;,
\end{equation}
the third metric is
\begin{equation}
\label{eq:fHM3}
\text{HM3:}\qquad f_{\rm HM3}(r)= 1-\frac{2M+\alpha j}{r} + \frac{2\alpha jM}{r^2} - \frac{\alpha M}{r^2}e^{-r/M} \left[ r-\alpha j e^{(r-\alpha j)/M} \right] \; ,
\end{equation}
with horizon radius, $r_{\rm h}=\alpha j\geq 2M$. Thus, HM3 depends explicitly on the parameter pair $(\alpha,j)$, with the combination $\alpha j$ governing the horizon structure and several associated observables.

\par Finally, for
\begin{equation}
Q^2=\alpha M(2M+\alpha j)e^{-(2M+\alpha j)/M} \; ,
\end{equation}
we obtain
\begin{equation}\label{eq:fHM4}
\text{HM4:}\qquad f_{\rm HM4}(r) = 1-\frac{2M+\alpha j}{r} - \frac{\alpha M}{r^2}e^{-r/M} \left[ r-(2M+\alpha j)e^{[r-(2M+\alpha j)]/M} \right] \; ,
\end{equation}
with $r_{\rm h}=2M+\alpha j\geq 2M.$
The Schwarzschild limit is recovered when the deformation sector is removed through $\alpha\rightarrow0$, provided the remaining parameters are scaled consistently.

\par Although HM2--HM4 exhibit an RN-like structure, the parameter $Q$ should not in general be interpreted as an electromagnetic charge. Instead, it serves as an effective charge associated with the additional source sector generated by the gravitational decoupling construction. In particular, these geometries originate from a Schwarzschild seed together with the corresponding deformation sector, rather than from an RN seed. The RN-like form is nevertheless useful for characterizing their parameter dependence and horizon structure, as discussed below.

\par An additional qualification is required for HM4. For the generic DEC parent geometry Eq.~\eqref{eq:fDEC}, the effective energy density in the $(-,+,+,+)$ convention adopted here is
\begin{equation}
 k^2\widetilde{\rho}\,(r)  =  \frac{Q^2}{r^4} - \frac{\alpha e^{-r/M}}{r^2} \;,
\end{equation}
where $k^2=8\pi G_N$ is the gravitational coupling introduced in Eq.~\eqref{eq:EFE}. For HM4, using
\begin{equation}
Q^2=\alpha M r_{\rm h} e^{-r_{\rm h}/M} \;, \qquad r_{\rm h}=2M+\alpha j \;, \nonumber
\end{equation} 
the effective energy density evaluated at the horizon becomes
\begin{equation}
 k^2\widetilde{\rho}(r_{\rm h})
 =
 \frac{\alpha e^{-r_{\rm h}/M}}{r_{\rm h}^3}
 \left(M-r_{\rm h}\right) \;.
\end{equation}
Since $r_{\rm h}\geq2M$, $k^2\widetilde{\rho}(r_{\rm h}) <0$ for all $\alpha>0$. Consequently, the nontrivial HM4 analytic subclass does not belong to the DEC-admissible sector of the parent family in Eq.~\eqref{eq:fDEC}. HM4 thereby serves in the analysis as an exploratory analytic deformation instead of a physically DEC-admissible branch.

\par The four metrics considered here thus provide a controlled set of static, spherically symmetric hairy black hole geometries that enable a systematic investigation of how distinct geometric deformations affect horizon properties, photon spheres, shadow observables, and QNM spectra, while also allowing us to distinguish physically admissible branches from exploratory configurations. Before turning to the optical and perturbative properties of these solutions, we first examine the parameter restrictions associated with the horizon structure, energy conditions, and comparison criterion adopted in this work.


\section{Geometric and observational properties of hairy black holes}\label{analysis of metric models}

We now examine the geometric and observational properties of the hairy black hole metrics introduced in the previous section. Our focus is on observables that can be compared directly with the Schwarzschild seed geometry, namely the horizon structure, photon sphere properties, shadow characteristics, EHT-scale angular observables, and the associated perturbative signatures discussed later through the QNM spectrum. Taken together, these quantities provide a complementary set of diagnostics with which to assess the phenomenological impact of the deformation parameters.


\subsection{Parameter space and physically-motivated restrictions}
\label{sec:parameter_space}

As discussed in Sec.~\ref{GD}, the deformation parameter $\alpha$ emerges naturally from the gravitational decoupling construction and controls the departure from the underlying seed geometry. Throughout this work, we restrict our attention to $\alpha \geq 0$. Additional restrictions are model dependent and arise from requiring that the spacetime possess a well-defined event horizon,\footnote{When multiple horizons are present, we also impose that the event horizon is identified with the largest real, positive root of $f(r)$.} the reality of any effective charge parameters, and (for the physically admissible branches) the corresponding energy-condition requirements. As discussed below, configurations that fall outside these controlled domains are retained only where explicitly identified as exploratory.

\par For HM1, the point $r_{\rm h}=2M$ is a horizon for all $\alpha$, but it remains the outer reference horizon only within a restricted range. Requiring that $f_{\rm HM1}(r)\geq0$ for $r\geq2M$ yields
\begin{equation}
0\leq \alpha_{\rm HM1}\leq e^2 \approx 7.3891\;.
\end{equation}
At the endpoint $\alpha=e^2$, the horizon at $r=2M$ becomes degenerate since $f'_{\rm HM1}(2M)=0$. Therefore, whenever a non-extremal horizon is required (for example in the construction of the tortoise coordinate used in the perturbation analysis), we adopt the stricter condition
\begin{equation}
0\leq \alpha_{\rm HM1}<e^2 \;.
\end{equation}
Values with $\alpha>e^2$ may still be examined as illustrative deformations; however, they no longer belong to the physical comparison branch associated with the reference horizon $r_{\rm h}=2M$ and should instead be interpreted using the corresponding outer event horizon of the geometry.

\par For HM2, the horizon is fixed at
\begin{equation}
r_{\rm h}^{\rm HM2}=2M \;,
\end{equation}
for the analytic subfamily defined by
\begin{equation}
Q^2=4\alpha\left(\frac{M}{e}\right)^2 \; , \qquad j = \frac{M}{e^2} \; .
\end{equation}
In this subclass, the parameters $Q$, $M$, and $\alpha$ are not independent. For fixed $M$, increasing $\alpha$ simultaneously increases the effective charge parameter $Q$.

The commonly used parametrization for charged black holes introduces the additional condition,\footnote{Ovalle \textit{et al.} \cite{Ovalle:2020kpd} discussed that the DEC family of solutions can be rewritten as black hole solutions from a theory of nonlinear electrodynamics coupled to gravity, characterized by a horizon structure similar to that of a RN solution,
\[
r_{\rm h} = r_{ \rm RN} \left( 1 - \alpha\frac{e^{-K}}{K}\right)^{-1} \geq r_{ \rm RN} \;,
\]
where $K=r_{\rm h}/M$. See Eqs. 103--106 of Ref. \cite{Ovalle:2020kpd}.}
\begin{equation}
1-\frac{\alpha}{2e^2}>0 \;,
\end{equation}
which gives
\begin{equation}
0\leq \alpha_{\rm HM2}<2e^2 \approx 14.7781\;.
\end{equation}
This restriction should be viewed primarily as preserving the RN-like interpretation of the solution branch rather than enforcing the existence of the horizon itself, which remains fixed at $r_{\rm h}=2M$ for the analytic subclass considered here.

\par For HM3, the horizon condition is
\begin{equation} \label{eq:rhHM3}
r_{\rm h}^{\rm HM3}=\alpha j\geq2M \;.
\end{equation}
Consequently, the admissible parameter space is naturally described by the pair $(\alpha,j)$ rather than by $\alpha$ alone. If one imposes the comparison criterion $r_{\rm h}^{\rm HM3}=2M$, then
\begin{equation}
j=\frac{2M}{\alpha} \; , \qquad \alpha>0 \; .
\end{equation}
Any scan in $\alpha$ performed under the fixed-horizon comparison criterion therefore induces a corresponding variation in $j$. 

\par The DEC restrictions of the parent family specified in Eq.~\eqref{eq:fDEC} further require that
\begin{equation}
j\geq\frac{M}{e^2} \;,
\qquad
Q^2\geq4\alpha\left(\frac{M}{e}\right)^2 .
\end{equation}
Combining the first of these conditions with $j=2M/\alpha$ gives
\begin{equation}
\frac{2M}{\alpha}\geq\frac{M}{e^2}
\qquad\ \Rightarrow \qquad
\alpha\leq2e^2 \;.
\end{equation}
The corresponding condition on $Q^2$ yields the same restriction. Thus, along the fixed-horizon comparison branch, the DEC-admissible parameter range for HM3 becomes
\begin{equation}
0<\alpha_{\rm HM3}\leq2e^2 \;.
\end{equation}
The benchmark values that we shall consider, $viz.$ $\alpha=1,2,3$, therefore all lie within this admissible range. We reiterate that since $j$ varies with $\alpha$ along this branch, HM3 benchmark points should be understood as the corresponding parameter pairs $(\alpha,j)$.

\par For HM4, the horizon condition is
\begin{equation} \label{eq:rhHM4}
r_{\rm h}^{\rm HM4}=2M+\alpha j\geq2M \; , \qquad \alpha j\geq0 \; .
\end{equation}
Imposing the same comparison criterion, $r_{\rm h}^{\rm HM4}=2M$, requires that $\alpha j=0$. For a nontrivial scan with $\alpha>0$, this selects
\begin{equation}
j=0 \;.
\end{equation}
However, as shown in Sec.~\ref{GD}, the resulting HM4 configurations with $j=0$ and $\alpha>0$ are not DEC-admissible. Substituting the HM4 charge relation into the effective energy density obtained from the Einstein field equations gives
\begin{equation}
k^2\widetilde{\rho}(r_{\rm h}) = \frac{\alpha e^{-r_{\rm h}/M}}{r_{\rm h}^3}
\left(M-r_{\rm h}\right) <0
\end{equation}
for every $\alpha>0$ and $r_{\rm h}\geq2M$. Moreover, this conclusion persists for $j>0$, since in this case the horizon is shifted to $r_{\rm h}>2M$ without changing the sign of the above expression. Relaxing the fixed-horizon criterion therefore does not restore DEC admissibility within this analytic HM4 subclass. We consequently treat all HM4 configurations with $\alpha>0$ considered below as exploratory geometric and perturbative benchmarks rather than as predictions from the DEC-admissible sector.

\par The fixed-horizon criterion therefore affects HM3 and HM4 in qualitatively different ways. For HM3, it imposes the relation $j=2M/\alpha$, while retaining the DEC-admissible range $0<\alpha\leq2e^2$. For HM4, by contrast, the same criterion requires $j=0$, while all configurations with $\alpha>0$ remain outside the DEC-admissible sector, irrespective of whether the horizon condition is fixed or relaxed. To illustrate the sensitivity of the perturbative observables to the horizon restriction itself, we nevertheless present additional QNM results in Tables~\ref{tab:wkb_qnm_models_2}--\ref{tab:qnm_wkb_iaim_comparison} for exploratory configurations with $r_{\rm h}>2M$.

In the numerical analysis that follows, we accordingly distinguish between parameter choices belonging to physically controlled comparison branches and those introduced solely for exploratory purposes. The latter are retained to illustrate how the optical and perturbative observables respond to geometric deformations beyond the physically admissible parameter domain.


\subsection{Horizon structure}

\begin{figure*}[h]
\centering
\captionsetup[subfigure]{justification=raggedleft,singlelinecheck=false}
\begin{subfigure}[b]{0.495\textwidth}
\includegraphics[width=\linewidth]{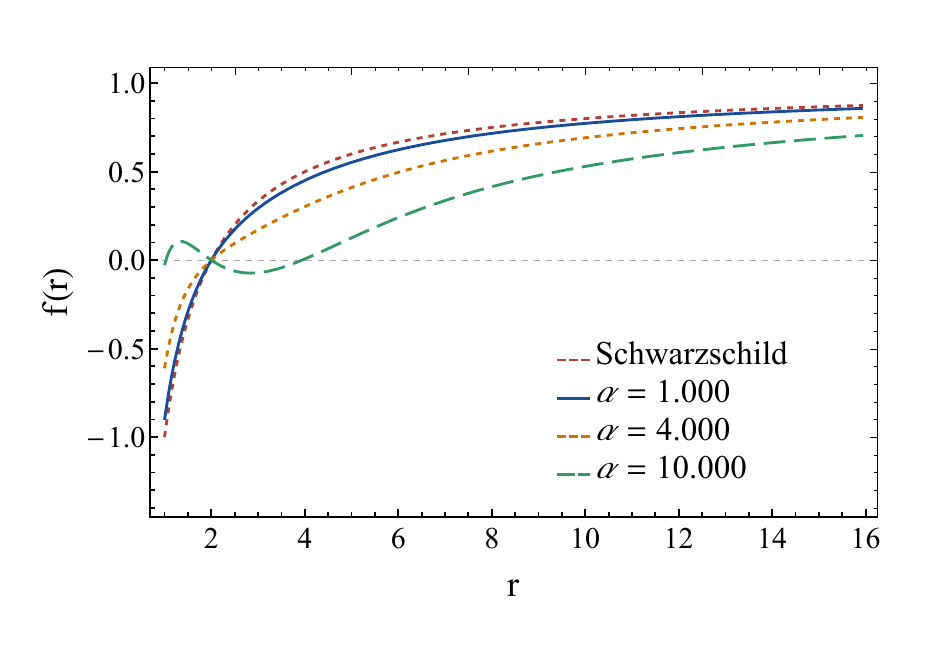}
\vspace{-1cm}
\caption{\it Hairy metric 1} \label{fig:horizon_hm1} \end{subfigure}
\hfill
\begin{subfigure}[b]{0.495\textwidth}
\centering
\includegraphics[width=\linewidth]{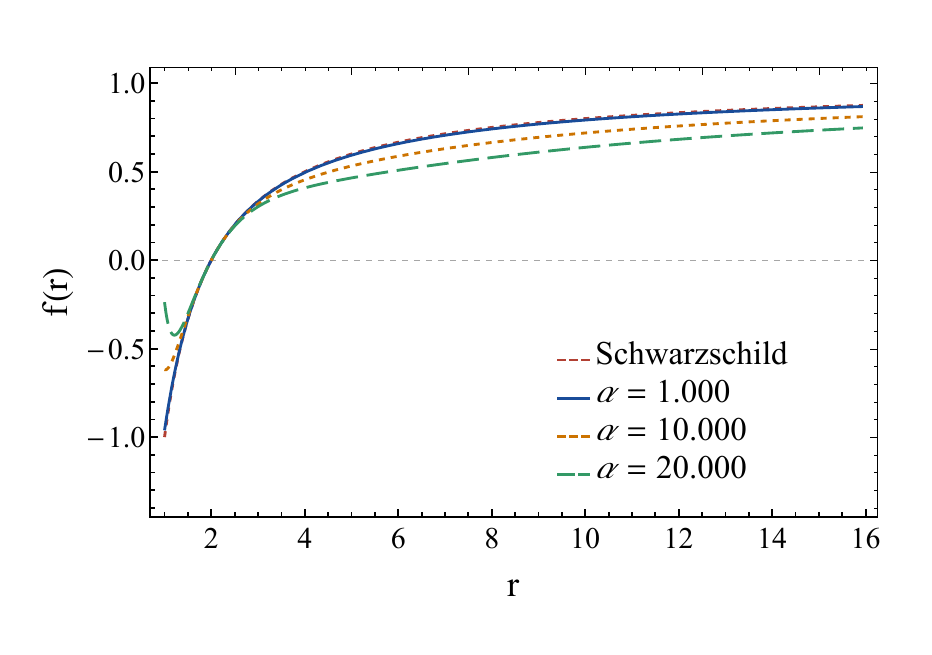}
\vspace{-1cm}
\caption{\it Hairy metric 2} \label{fig:horizon_hm2}
\end{subfigure}
\vspace{0.15cm}
\begin{subfigure}[b]{0.495\textwidth}
\centering
\includegraphics[width=\linewidth]{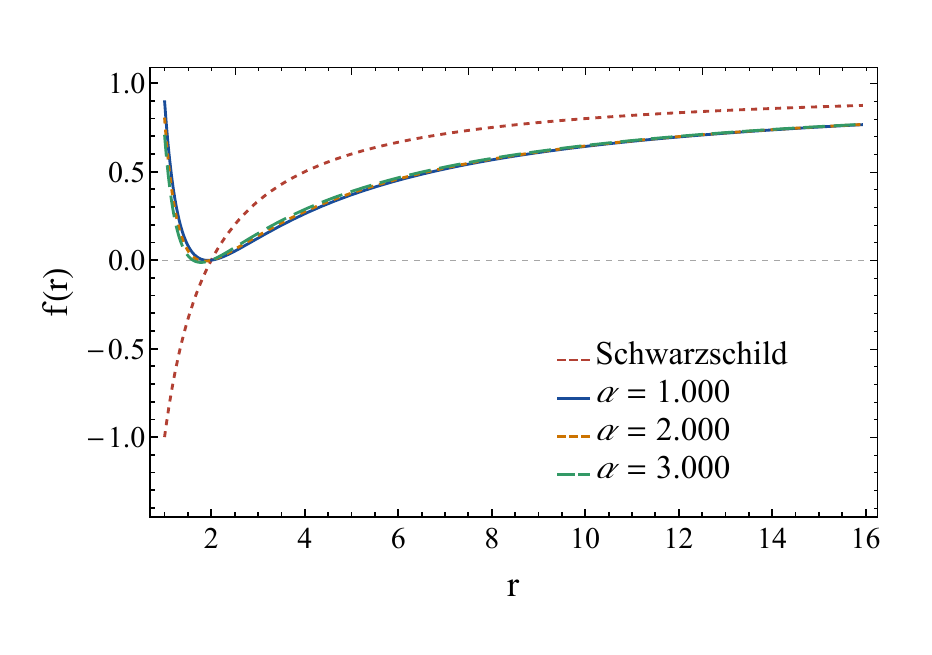}
\vspace{-1cm}
\caption{\it Hairy metric 3} \label{fig:horizon_hm3}
\end{subfigure}
\hfill
\begin{subfigure}[b]{0.495\textwidth}
\centering
\includegraphics[width=\linewidth]{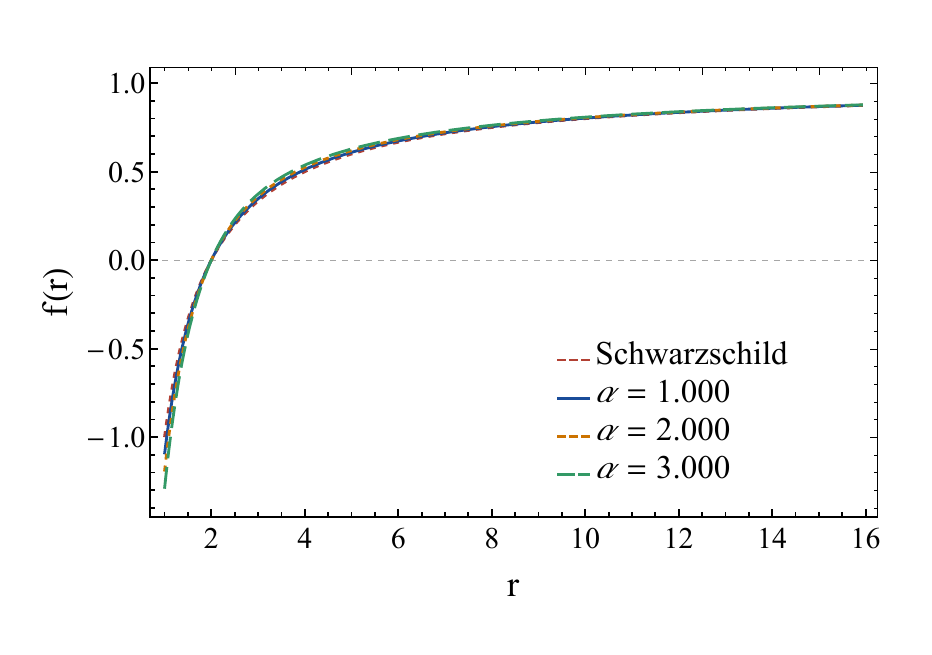}
\vspace{-1cm}
\caption{\it Hairy metric 4} \label{fig:horizon_hm4}
\end{subfigure}
\vskip 0.25cm
\caption{\textit{Metric functions $f(r)$ for HM1--HM4, with the Schwarzschild solution shown as a reference. The radial coordinate is expressed in units of the seed mass, with $M=1$. For HM1 and HM2, the curves illustrate the dependence on the deformation parameter $\alpha$. For HM3 and HM4, the comparison is performed subject to the fixed-horizon condition $r_{\rm h}=2M$. This requires $j=2M/\alpha$ for HM3, so that the displayed values $\alpha=1,2,3$ correspond to $j=2,1,2/3$, respectively, while for HM4 with $\alpha>0$ the same condition requires $j=0$. Parameter choices outside the controlled domains identified in Sec.~\ref{sec:parameter_space}, including $\alpha=10$ for HM1, $\alpha=20$ for HM2, and all $\alpha>0$ HM4 configurations, and are shown only for exploratory comparison.}}
\label{fig:horizon_plot}
\end{figure*}

We now examine how the horizon structure of the four hairy black hole geometries respond to variations of the deformation parameters. For the static, spherically symmetric metrics considered here, horizons are determined by the positive roots of $f(r)=0$, with the event horizon identified as the outermost such root. For HM1, $r_{\rm h}=2M$ remains the outer horizon within the controlled range $0\leq\alpha<e^2$. At the endpoint $\alpha=e^2$, this horizon becomes degenerate; for $\alpha>e^2$, additional roots may appear and the outer horizon is shifted away from $r=2M$. Thus, values beyond this range should be interpreted as exploratory deformations outside the fixed-horizon branch. 

For HM2, the analytic subfamily considered here preserves $r_{\rm h}=2M$ by construction, while the additional restriction on $\alpha$ is associated with maintaining the RN-like interpretation of the branch. 

For HM3 and HM4, the horizon depends explicitly on both $\alpha$ and $j$, with $r_{\rm h}^{\rm HM3}=\alpha j$ and $r_{\rm h}^{\rm HM4}=2M+\alpha j$. Consequently, any fixed-horizon comparison must specify the corresponding relation between $\alpha$ and $j$: imposing $r_{\rm h}=2M$ requires $j=2M/\alpha$ for HM3, while for HM4 it selects the slice $\alpha j=0$.

Fig.~\ref{fig:horizon_plot} shows the metric functions $f(r)$ for the four hairy black hole geometries, with the Schwarzschild profile included as a reference. For HM1, the deformation becomes increasingly pronounced as $\alpha$ is increased. The value $\alpha=10$ is included only for illustrative purposes, since it lies beyond the controlled range and demonstrates the change in global horizon structure associated with the appearance of additional roots. For HM2, increasing $\alpha$ modifies the exterior profile while leaving the reference horizon unchanged. Larger deformation values suppress $f(r)$ in the near-horizon and intermediate radial regions and lead to a slower approach to the asymptotically flat limit, reflecting the increasing contribution of the effective charge sector; values beyond $\alpha=2e^2$ should therefore be regarded as illustrative rather than as part of the controlled charged branch.

The lower panels illustrate the different consequences of imposing $r_{\rm h}=2M$ on HM3 and HM4. For HM3, this requires $\alpha j =2M$, so that varying $\alpha$ simultaneously changes $j$. The exterior metric nevertheless remains substantially different from the Schwarzschild profile, demonstrating that fixing the horizon radius does not eliminate the geometric effects of the deformation, influencing photon rings, shadow observables, and perturbative dynamics. For HM4, the same criterion requires that $j=0$ for $\alpha>0$; the plotted curves correspond to this choice of $j=0$. The resulting metric functions remain comparatively close to the Schwarzschild profile in the exterior region, with mild deviations in the near-horizon and intermediate regions that persist away from the horizon. However, as established in Sec.~\ref{sec:parameter_space}, these HM4 configurations are exploratory rather than DEC-admissible; their inclusion allows us to examine how the corresponding geometric deformation propagates into the optical and perturbative observables considered below.


\subsection{Photon spheres and shadow observables}
\label{photon_shadow}

\par As discussed in the Introduction, unstable circular photon orbits provide a useful probe of the black hole geometry. In a static, spherically symmetric spacetime, the photon sphere radius determines the location of the relevant circular null orbit, while the associated critical impact parameter defines the boundary between null rays that escape to infinity and those captured by the black hole. The latter therefore sets the characteristic geometric scale of the black hole shadow. The properties of the photon sphere are also related, in the eikonal regime, to the characteristic oscillation and damping scales of QNMs, providing a useful connection between the null geodesic structure and the perturbative properties discussed in Sec.~\ref{QNMS}.

\par The photon sphere radius is determined by
\begin{equation}
r_{\rm ph}f'(r_{\rm ph})-2f(r_{\rm ph})=0 \;,
\qquad
f'(r)=\frac{df(r)}{dr} \;.
\end{equation}
\noindent Among the positive solutions of this equation, we identify
$r_{\rm ph}$ with the outer circular null orbit lying outside the event
horizon and satisfying the instability condition,
\begin{equation}
\frac{d^2}{dr^2}
\left(\frac{f(r)}{r^2}\right)
\bigg \vert_{r=r_{\rm ph}}<0 \;.
\end{equation}

\par For the hairy metrics considered here, $r_{\rm ph}$ depends on the corresponding metric parameters, which we denote schematically by
\begin{equation}
r_{\rm ph}=r_{\rm ph}(M,\alpha,j) \;.
\end{equation}
The associated critical impact parameter is
\begin{equation}
b_{\rm ph} = \frac{r_{\rm ph}}{\sqrt{f(r_{\rm ph})}}
\;.
\end{equation}

\par The numerical results presented in this subsection are expressed in units of the seed mass, with $M=1$. For HM3 and HM4, $j$ is not varied independently. Instead, the horizon comparison criterion $r_{\rm h}=2M$ is imposed directly. For HM3, this requires
\begin{equation}
j=\frac{2M}{\alpha} \;,
\end{equation}
so that the benchmark values $\alpha=1,2,3$ correspond, respectively, to
\begin{equation}
(\alpha,j)=(1,2),\quad(2,1),\quad(3,2/3)
\end{equation}
for $M=1$. For HM4, the same horizon condition requires $j=0$ for every $\alpha>0$. As established in Sec.~\ref{sec:parameter_space}, the HM3 benchmark configurations lie within the DEC-admissible range, whereas all $\alpha>0$ HM4 configurations considered here are retained solely as exploratory geometric benchmarks.

\par It is important to distinguish the seed-mass normalization used here from a comparison at fixed asymptotic mass. Since the ADM mass generally differs from $M$ and depends on the deformation parameters, the values of $b_{\rm ph}$ quoted below should first be interpreted as geometric quantities in units of the seed mass. The corresponding ADM-normalized critical impact parameters, required for comparison at fixed observed mass, are introduced in Sec.~\ref{eht_comparison}.

\par
\begin{figure*}
\centering
\captionsetup[subfigure]{justification=raggedleft,singlelinecheck=false}
\begin{subfigure}[b]{0.45\linewidth}
\centering
\includegraphics[width=\linewidth]{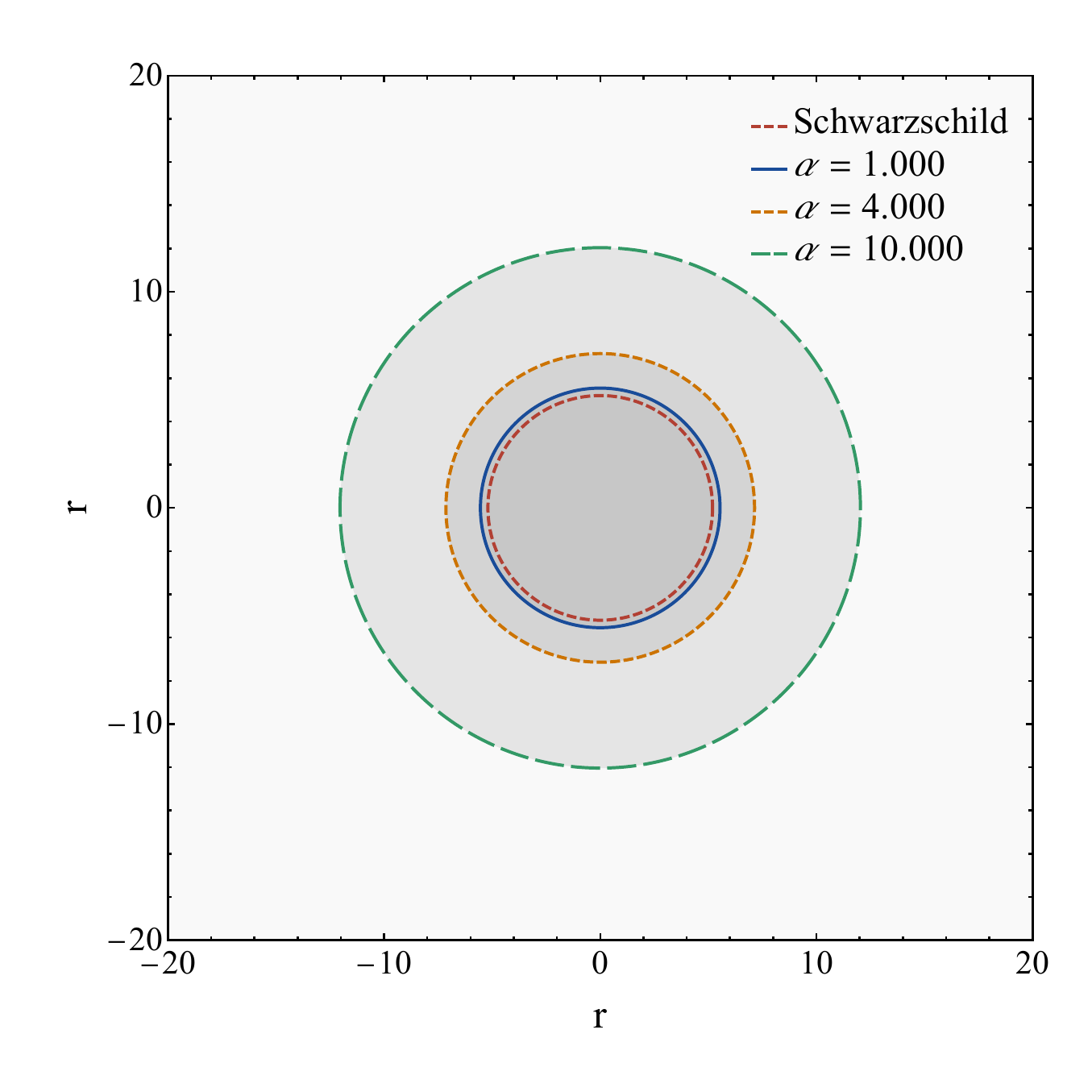}
\vspace{-1cm}
\caption{\it Hairy metric 1}
\label{fig:shadow_hm1}
\end{subfigure}
\hfill
\begin{subfigure}[b]{0.45\textwidth}
\centering
\includegraphics[width=\textwidth]{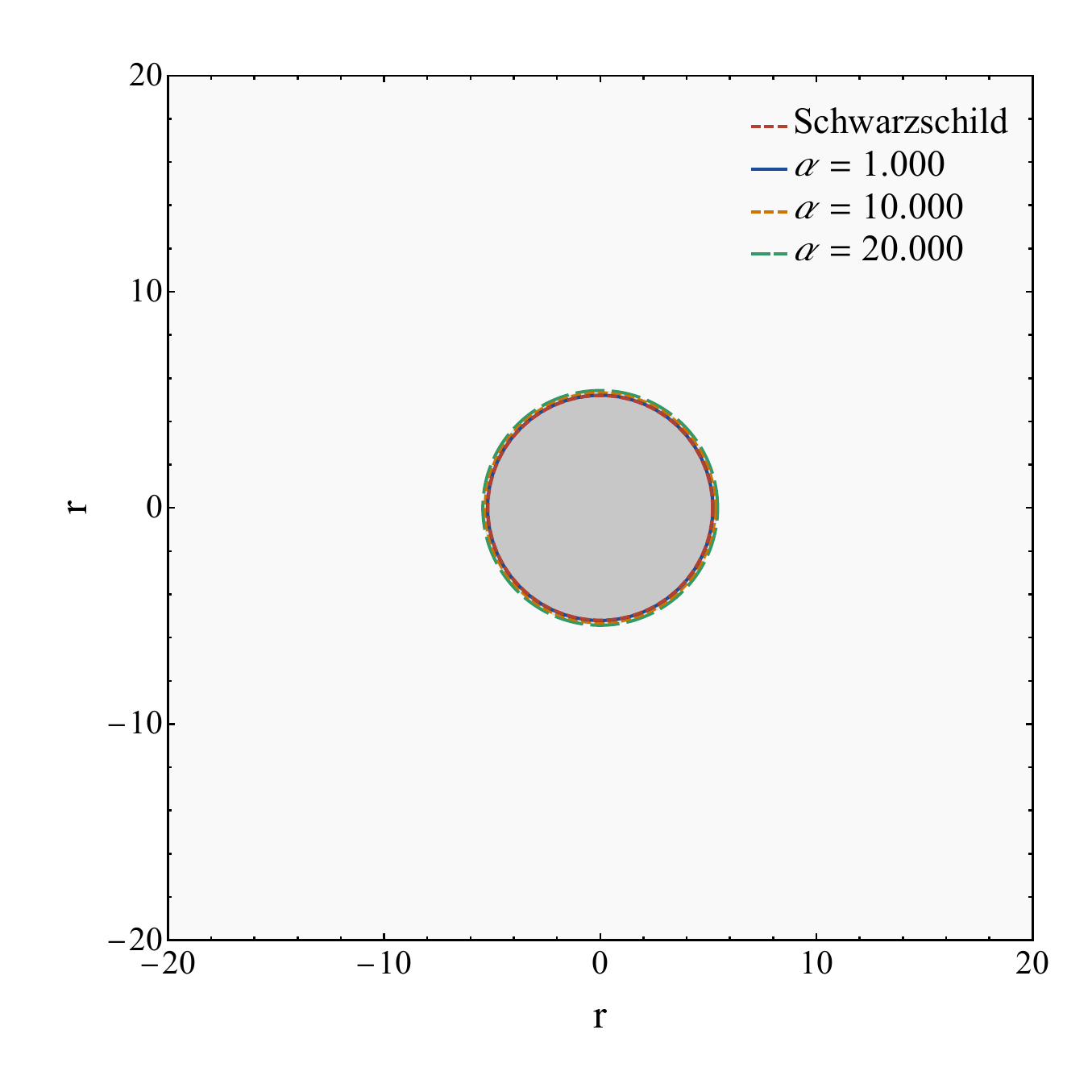}
\vspace{-1cm}
\caption{\it Hairy metric 2}
\label{fig:shadow_hm2}
\end{subfigure}
\vspace{0.05cm}
\begin{subfigure}[b]{0.45\textwidth}
\centering
\includegraphics[width=\textwidth]{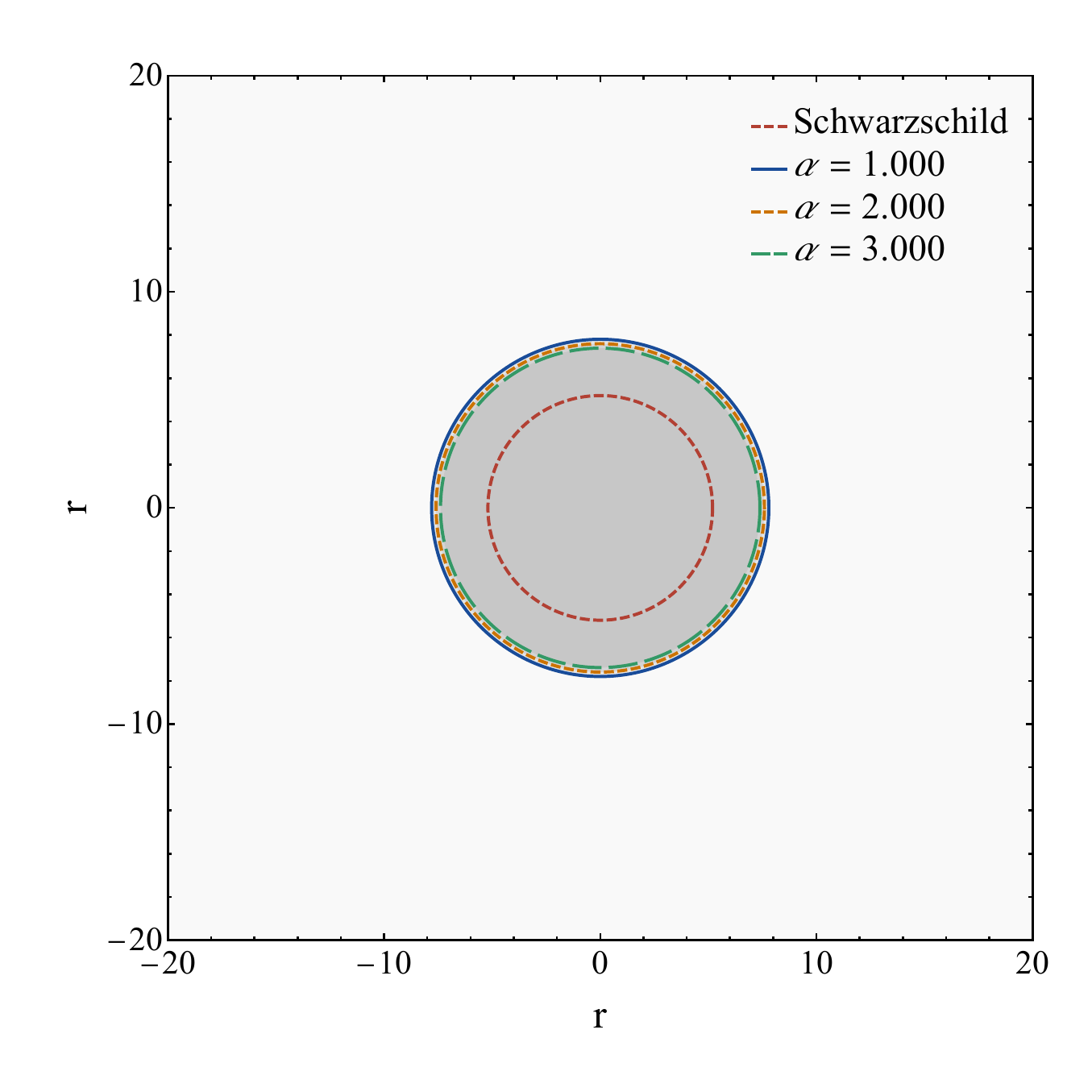}
\vspace{-1cm}
\caption{\it Hairy metric 3}
\label{fig:shadow_hm3}
\end{subfigure}
\hfill
\begin{subfigure}[b]{0.45\textwidth}
\centering
\includegraphics[width=\textwidth]{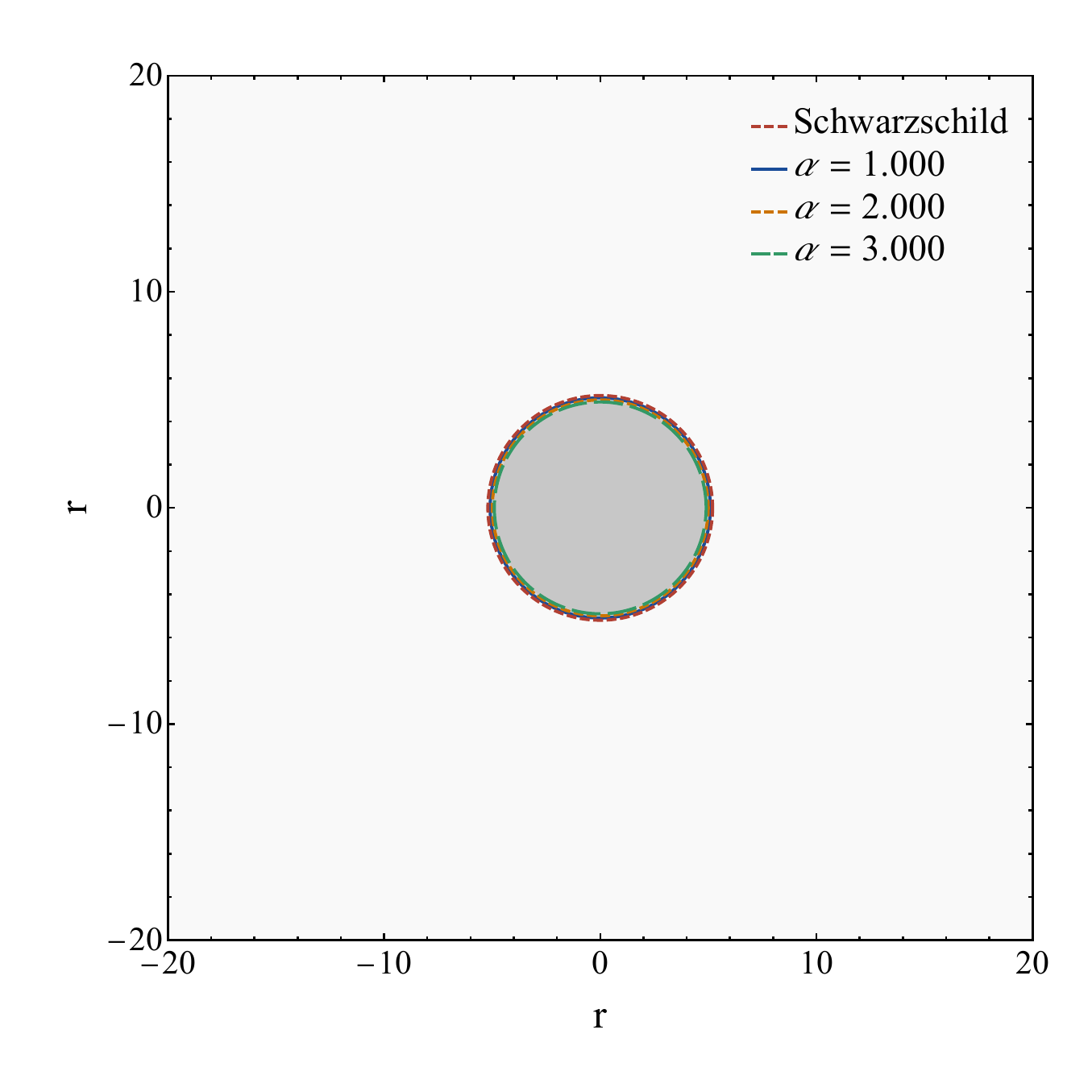}
\vspace{-1cm}
\caption{\it Hairy metric 4}
\label{fig:shadow_hm4}
\end{subfigure}
\vskip 0.25cm
\caption{\textit{Critical curves associated with the critical impact parameter $b_{\rm ph}$ for the four hairy metrics, shown in units of the seed mass $M=1$. The Schwarzschild critical curve is included as a common reference. For HM3, $r_{\rm h}=2M$ is maintained through $j=2M/\alpha$, while for HM4 the same condition requires $j=0$ for $\alpha>0$. Parameter choices outside the controlled domains identified in Sec.~\ref{sec:parameter_space}, including all nonzero-$\alpha$ HM4 configurations, are shown only for exploratory comparison.}}
\label{fig:BH_shadow_plots}
\end{figure*}

 The resulting photon sphere radii and critical impact parameters are summarized in Table~\ref{tab:photon_rings_and_BH_Shadows}, while the corresponding critical curves are shown in Fig.~\ref{fig:BH_shadow_plots}. For all configurations displayed, the hierarchy
\begin{equation}
r_{\rm h}<r_{\rm ph}
\end{equation}
is preserved. The response to increasing $\alpha$ is nevertheless strongly metric dependent. In the seed-mass normalization adopted here, HM1 and HM2 exhibit an increase in $b_{\rm ph}$ over the sampled values of $\alpha$, whereas HM3 and the exploratory HM4 configurations show a gradual decrease. HM3 nevertheless retains substantially larger values of both $r_{\rm ph}$ and $b_{\rm ph}$ than Schwarzschild in this normalization throughout the benchmark range.

\begin{table*}
\centering
\caption{\textit{Photon sphere radii and critical impact parameters for the hairy metrics, in units of the seed mass $M=1$. \enquote{Sc.} denotes the Schwarzschild reference. For HM3, the fixed-horizon condition $r_{\rm h}=2M$ requires $j=2M/\alpha$, so that $\alpha=1,2,3$ correspond to $j=2,1,2/3$, respectively. For HM4, $r_{\rm h}=2M$ with $\alpha>0$ requires $j=0$; these configurations are exploratory and are not DEC-admissible. The HM1 configuration with $\alpha=10$ and the HM2 configuration with $\alpha=20$ are likewise shown as exploratory extensions beyond the controlled parameter domains identified in Sec.~\ref{sec:parameter_space}.}}
\label{tab:photon_rings_and_BH_Shadows}
\vskip 0.2cm
\renewcommand{\arraystretch}{1.8} \setlength{\tabcolsep}{3pt} 
\begin{tabular}{|c|ccc|c|ccc|c|ccc|c|ccc|} \hline \multicolumn{4}{|c|}{\text{Hairy Metric 1}} & \multicolumn{4}{c|}{\text{Hairy Metric 2}} & \multicolumn{4}{c|}{\text{Hairy Metric 3}} & \multicolumn{4}{c|}{\text{Hairy Metric 4}} \\ \hline $\alpha$ & $r_{\rm h}$ & $r_{\rm ph}$ & $b_{\rm ph}$ & $\alpha$ & $r_{\rm h}$ & $r_{\rm ph}$ & $b_{\rm ph}$ & $\alpha$ & $r_{\rm h}$ & $r_{\rm ph}$ & $b_{\rm ph}$ & $\alpha$ & $r_{\rm h}$ & $r_{\rm ph}$ & $b_{\rm ph}$ \\ \hline \text{Sc.} & 2.0000 & 3.0000 & 5.1962 & \text{Sc.} & 2.0000 & 3.0000 & 5.1962 & \text{Sc.} & 2.0000 & 3.0000 & 5.1962 & \text{Sc.} & 2.0000 & 3.0000 & 5.1962 \\ 1 & 2.0000 & 3.0395 & 5.5419 & 1 & 2.0000 & 2.9915 & 5.2083 & 1 & 2.0000 & 3.8589 & 7.7999 & 1 & 2.0000 & 2.9708 & 5.0934 \\ 4 & 2.0000 & 3.3960 & 7.1389 & 10 & 2.000 & 2.9187 & 5.3154 & 2 & 2.0000 & 3.7226 & 7.5955 & 2 & 2.0000 & 2.9451 & 4.9953 \\ 10 & 3.9410 & 6.7018 & 12.0371 & 20 & 2.0000 & 2.8466 & 5.4294 & 3 & 2.0000 & 3.5942 & 7.3883 & 3 & 2.000 & 2.9223 & 4.9016\\ \bottomrule \end{tabular} 
\end{table*}

\par Table~\ref{tab:photon_rings_and_BH_Shadows} demonstrates that the geometric response to the deformation is not universal. For HM1, increasing $\alpha$ initially shifts the photon sphere outwards while increasing the critical impact parameter. The $\alpha=10$ configuration lies beyond the controlled fixed-horizon domain: its outer event horizon has moved to $r_{\rm h} \simeq 3.9410$, accompanied by a substantial increase in both $r_{\rm ph}$ and $b_{\rm ph}$. This point is therefore retained only to illustrate the continuation of the geometric deformation beyond the controlled HM1 branch.

\par For HM2, the horizon remains at $r_{\rm h}=2M$ throughout the sampled range. Increasing $\alpha$ shifts the photon sphere slightly inwards, while $b_{\rm ph}$ increases mildly. This illustrates that the critical impact parameter is not determined solely by $r_{\rm ph}$, but also depends on the metric function evaluated at the photon sphere through
\begin{equation}
b_{\rm ph} = \frac{r_{\rm ph}}{\sqrt{f(r_{\rm ph})}} \;.
\end{equation}
The $\alpha=1$ and $\alpha=10$ configurations lie within the adopted controlled comparison range, whereas $\alpha=20$ is retained only as an exploratory extension.

\par For HM3, the fixed-horizon condition maintains $r_{\rm h}=2M$ while varying $\alpha$ requires a compensating change in $j$. Across the DEC-admissible benchmark values $\alpha=1,2,3$, both $r_{\rm ph}$ and $b_{\rm ph}$ decrease gradually with increasing $\alpha$, although they remain substantially larger than their Schwarzschild values when expressed in units of the seed mass. As emphasized above, this statement concerns the seed-mass normalization; the comparison changes once the corresponding ADM mass is taken into account in Sec.~\ref{eht_comparison}.

\par For the exploratory HM4 configurations, the fixed-horizon condition requires $j=0$. Both $r_{\rm ph}$ and $b_{\rm ph}$ decrease monotonically with increasing $\alpha$, giving critical curves that are slightly smaller than the Schwarzschild reference in seed-mass units. Since these nonzero-$\alpha$ configurations do not belong to the DEC-admissible sector, these results should be interpreted as geometric properties of the HM4 deformation rather than as predictions from a physically controlled DEC branch.

\par The photon sphere radius and critical impact parameter therefore provide complementary geometric diagnostics of the four deformations. The former determines the location of the relevant circular null orbit, while the latter characterizes the capture boundary seen by a distant observer. Their metric-dependent trends also provide useful geometric intuition for the QNM behavior considered in Sec.~\ref{QNMS}. There, however, the quantitative QNM spectra are calculated independently rather than inferred from the eikonal correspondence.

\subsection{Critical angular diameter and comparison with EHT observations}
\label{eht_comparison}

In the preceding subsections, we computed the horizon radius, photon radius, and shadow radius for each of the hairy metrics in natural units, with the seed mass set to $M=1$. Thus far, we have adopted this convention for theoretical comparison. However, when comparing with actual observed data, we need to reinstate the relevant normalization. Here, following the geometric shadow and lensing-motivated analyses of Refs.~\cite{Crisnejo:2019ril,mustafa2024testing}, we compare the shadow angular diameter predicted by each hairy metric against corresponding observed values obtained via the EHT. Specifically, we use the critical impact parameter to estimate the angular radius of the photon sphere critical curve, $\theta_\infty=b_{\rm ph}^{\rm phys}/D_{\rm OL}$, and hence the corresponding critical angular diameter. However, when comparing against astrophysical observations, we must re-introduce SI units and identify the appropriate mass scale inferred by a distant observer.

For a hairy black hole spacetime, the mass parameter appearing in the metric need not coincide with the asymptotic mass measured at infinity. In particular, if the ADM mass differs from the seed mass according to
\begin{equation}
M_{\rm ADM} = M_{\rm ADM}(\alpha) \;,
\end{equation}
then the appropriate dimensionless quantity is the impact parameter normalized by the ADM mass,
\begin{equation}
\bar b_{\rm ph} = \frac{b_{\rm ph}}{M_{\rm ADM}} \;.
\end{equation}
The ADM mass for each hairy metric is displayed in Table~\ref{tab:adm_masses}. The corresponding physical critical impact parameter for a black hole of observed mass $M_{\rm obs}$ is
\begin{equation}
b_{\rm ph}^{\rm phys} = \bar b_{\rm ph} \frac{G_N M_{\rm obs}}{c^2} \; .
\end{equation}

For the fixed-horizon configurations used for HM3 and HM4 in Secs.~\ref{sec:parameter_space} and~\ref{photon_shadow}, these expressions simplify further. For HM3, $j=2M/\alpha$ implies
\begin{equation}
M_{\rm ADM}^{\rm HM3}=2M \;,
\end{equation}
independently of $\alpha$ along the comparison branch. For HM4, $j=0$ gives
\begin{equation}
M_{\rm ADM}^{\rm HM4}=M \;.
\end{equation}
This distinction is important: the comparatively large values of $b_{\rm ph}$ found for HM3 in seed-mass units do not translate directly into a correspondingly large critical angular diameter when configurations are compared at fixed observed mass.

\begin{table*} 
\centering 
\caption{\textit{Large-$r$ asymptotic expansion of the metric function $f(r)$ and corresponding ADM mass for HM1--HM4. The ADM mass represents the total asymptotic mass measured by a distant observer and therefore sets the appropriate physical mass scale for comparing the different geometries at fixed observed mass. It is read from the coefficient of the $1/r$ term and generally differs from the seed-mass parameter $M$ when the deformation parameters are nonzero.}} \label{tab:adm_masses} 
\vskip 0.2cm 
\renewcommand{\arraystretch}{2.2} 
\setlength{\tabcolsep}{10pt} 
\begin{tabular}{|c|c|c|} 
\hline 
\text{Metric} & \text{Asymptotic form of the metric function} & \text{ADM mass} \\ 
\hline 
HM1 & $\displaystyle f(r)\sim 1-\frac{2M}{r} \left(1 + \frac{\alpha}{e^2}\right)$ & $\displaystyle M_{\rm ADM} = M\left(1+\frac{\alpha}{e^2}\right)$ \\[5pt] 
\hline 
HM2 & $\displaystyle f(r)\sim 1-\frac{2M}{r}\left(1+\frac{\alpha}{2e^2}\right)$ & $\displaystyle M_{\rm ADM} = M\left(1+\frac{\alpha}{2e^2}\right)$ \\[5pt] 
\hline 
HM3 & $\displaystyle f(r)\sim 1-\frac{2M+\alpha j}{r}+O(r^{-2})$ & $\displaystyle M_{\rm ADM} = M+\frac{\alpha j}{2}$ \\[5pt] 
\hline 
HM4 & $\displaystyle f(r)\sim 1-\frac{2M+\alpha j}{r}+O(r^{-2})$ & $\displaystyle M_{\rm ADM} = M+\frac{\alpha j}{2}$ \\[5pt] 
\hline 
\end{tabular} 
\end{table*} 

\noindent The angular radius of the critical curve is therefore
\begin{equation}
\theta_{\infty} = \frac{b_{\rm ph}^{\rm phys}}{D_{\rm OL}} \; ,
\end{equation}
and the critical angular diameter is
\begin{equation}
d_{\rm metric} = 2\theta_{\infty} = 2\bar b_{\rm ph} \frac{G_N M_{\rm obs}}{c^2D_{\rm OL}} \;,
\label{eq:dmetric}
\end{equation}
where $D_{\rm OL}$ is the distance between the observer (O) and the lens (L). This quantity corresponds to the angular diameter of the photon sphere critical curve. Although closely related to the characteristic angular scale of a black hole image, $d_{\rm metric}$ should not be identified directly with the observed emission ring diameter. The latter also depends on the emitting plasma, source morphology, viewing geometry, calibration, and the mass-distance inference. We therefore use the EHT measurements below only to provide an illustrative comparison of angular scales, rather than to derive statistical constraints on the deformation parameters.

\par For the reference mass-distance scales adopted here, we use for M87*,
\begin{equation}
\frac{G_N M_{\rm M87*}}{c^2D_{\rm M87*}} \simeq 3.8 \,\mu{\rm as} \;,
\end{equation}
and for Sgr~A*,
\begin{equation}
\frac{G_N M_{\rm SgrA*}}{c^2D_{\rm SgrA*}} \simeq 4.8\,\mu{\rm as} \;.
\end{equation}
Thus,
\begin{equation}
d_{\rm metric}^{\rm M87*}(\alpha) = 2\bar b_{\rm ph}(\alpha) (3.8 \,\mu{\rm as}) \;,
\end{equation}
and
\begin{equation}
d_{\rm metric}^{\rm SgrA*}(\alpha) = 2\bar b_{\rm ph}(\alpha) (4.8 \,\mu{\rm as}) \;.
\end{equation}
We compare our predictions with the central-value EHT measurements of the angular diameter of the emission ring,
\begin{equation}
d_{\rm obs}^{\rm M87*} \approx 42 \,\mu{\rm as} \;,
\end{equation}
and
\begin{equation}
d_{\rm obs}^{\rm SgrA*} \approx 51.8 \,\mu{\rm as} \;.
\end{equation}
Here, $\mu{\rm as}$ denotes microarcseconds, where $1\mu{\rm as}=10^{-6}$ arcseconds; this is the natural angular unit for comparing the predicted shadow diameter with the angular scales resolved by the EHT. We reiterate that these observed emission ring diameters and the geometrically defined quantities $d_{\rm metric}$ are not identical observables; their comparison is used here solely to characterize the magnitude of the angular-scale shifts induced by the different metric deformations.

\par To quantify this illustrative comparison, we define the percentage difference between the predicted critical angular diameter and the observed emission ring central value for each metric, 
\begin{equation}
\Delta_{\rm metric}^{\rm M87*}(\alpha) = 100 \times \left| \frac{ d_{\rm metric}^{\rm M87*}(\alpha) - d_{\rm obs}^{\rm M87*} }{ d_{\rm obs}^{\rm M87*} } \right| \;,
\end{equation}
and
\begin{equation}
\Delta_{\rm metric}^{\rm SgrA*}(\alpha) = 100 \times \left| \frac{ d_{\rm metric}^{\rm SgrA*}(\alpha) - d_{\rm obs}^{\rm SgrA*} }{ d_{\rm obs}^{\rm SgrA*} } \right| \;.
\end{equation}
Here, $d_{\rm metric}^{\rm M87*}$ and $d_{\rm metric}^{\rm SgrA*}$ denote the critical angular diameters predicted for hairy metrics with the same observed mass and distance as M87* and Sgr~A*, respectively, but described by the corresponding hairy geometry as the hair parameter $\alpha$ is varied. Note carefully, however, that these percentage quantities are descriptive central-value differences only. They are not confidence levels, exclusion significances, or statistical measures of compatibility with the EHT data. The summary of comparisons between the hairy metrics HM1-HM4 and the observed M87* and Sgr~A* are displayed in Tables~\ref{tab:SGL_metric1}-\ref{tab:SGL_metric4}, respectively.

\par We reiterate that these fractional deviations provide a simple diagnostic of how strongly a given hairy metric departs from the observed angular scale. However, they should not by themselves be interpreted as exclusion criteria. A statistically meaningful test of the deformation parameters would require a consistent treatment of the observational uncertainties, the independently inferred mass and distance, and the modeling and calibration uncertainties connecting the geometrical critical curve to the observed emission structure. Such an analysis is beyond the scope of the present work. We therefore restrict the discussion to the descriptive central-value comparison defined above and do not infer observational bounds on $\alpha$ from these quantities.

\begin{table*}
\centering
\caption{\textit{ADM-normalized critical impact parameters and the corresponding critical angular diameters for HM1, together with their percentage differences from the adopted M87* and Sgr~A* emission ring central values. The $\alpha=$10 configuration lies outside the controlled domain identified in Sec.~\ref{sec:parameter_space}.}} \label{tab:SGL_metric1}
\vskip 0.2cm
\renewcommand{\arraystretch}{1.8} 
\setlength{\tabcolsep}{10pt}
\begin{tabular}{|c|c|c|c|c|c|c|}
\hline
$\alpha$ & $M_{\rm ADM}$ & $\bar{b}_{\rm ph}$ & $\displaystyle d_{\rm{HM1}}^{\text{M87*}}$ $(\mu\mathrm{as})$& $\displaystyle \Delta_{\rm{HM1}}^{\text{M87*}}$ $(\%)$& $\displaystyle d_{\rm{HM1}}^{\text{SgrA*}}$ $(\mu\mathrm{as})$ & $\displaystyle \Delta_{\rm{HM1}}^{\text{SgrA*}}$ $(\%)$ \\
\hline
\text{Sc.} & $1$ & $5.1962$ & $39.4908$& $5.9744$ & $49.8831$ & $3.7007$ \\
\hline
1 & 1.1353 & 4.8813 & 37.0979 & 11.6717 & 46.8605 & 9.5357 \\
\hline
$4$ & $1.5413$ & $4.6317$ & $35.2007$ & $16.1889$ & $44.4640$ & $14.1622$ \\ 
\hline $10$ & $2.3534$ & $5.1149$ & $38.8731$ & $7.4449$ & $49.1029$ & $5.2067$ \\
\hline
\end{tabular}
\vskip 0.5cm
\centering 
\caption{\textit{ADM-normalized critical impact parameters and the corresponding critical angular diameters for HM2, together with their percentage differences from the adopted M87* and Sgr~A* emission ring central values. The configurations $\alpha=$1 and 10 lie within the adopted controlled comparison range identified in Sec.~\ref{sec:parameter_space}, while $\alpha=$20 is shown only as an exploratory extension.
}} \label{tab:SGL_metric2} 
\vskip 0.2cm 
\renewcommand{\arraystretch}{1.8} 
\setlength{\tabcolsep}{10pt} 
\begin{tabular}{|c|c|c|c|c|c|c|} 
\hline 
$\alpha$ & $M_{\rm ADM}$ & $\bar{b}_{\rm ph}$ & $\displaystyle d_{\rm{HM2}}^{\text{M87*}}$ $(\mu\mathrm{as})$ & $\displaystyle \Delta_{\rm{HM2}}^{\text{M87*}}$ $(\%)$ & $\displaystyle d_{\rm{HM2}}^{\text{SgrA*}}$ $(\mu\mathrm{as})$ & $\displaystyle \Delta_{\rm{HM2}}^{\text{SgrA*}}$ $(\%)$ \\ 
\hline 
\text{Sc.} & $1$ & $5.1962$ & $39.4908$& $5.9744$ & $49.8831$ & $3.7007$ \\ 
\hline 
1 & 1.0677 & 4.8782 & 37.0742 & 11.7281 & 46.8306 & 9.5935 \\ 
\hline 
$10$ & $1.6767$ & $3.1702$ & $24.0933$ & $42.6349$ & $30.4337$ & $41.2477$ \\ 
\hline 
$20$ & $2.3534$ & $2.3071$ & $17.5338$ & $58.2528$ & $22.148$ & $57.2433$ \\ 
\hline 
\end{tabular} 
\vskip 0.5cm 
\centering 
\caption{\textit{ADM-normalized critical impact parameters and the corresponding critical angular diameters for HM3, together with their percentage differences from the adopted M87* and Sgr~A* emission ring central values. The fixed-horizon condition $r_{\rm h}=2M$ implies $j=2M/\alpha$ and hence $M_{\rm ADM}=2M$ for all three benchmark configurations. The values $\alpha=$1,2,3 lie within the DEC-admissible range identified in Sec.~\ref{sec:parameter_space}.
}} \label{tab:SGL_metric3} 
\vskip 0.2cm 
\renewcommand{\arraystretch}{1.8} 
\setlength{\tabcolsep}{10pt} 
\begin{tabular}{|c|c|c|c|c|c|c|} 
\hline 
$\alpha$ & $M_{\rm ADM}$ & $\bar{b}_{\rm ph}$ & $\displaystyle d_{\rm{HM3}}^{\text{M87*}}$ $(\mu\mathrm{as})$ & $\displaystyle \Delta_{\rm{HM3}}^{\text{M87*}}$ $(\%)$ & $\displaystyle d_{\rm{HM3}}^{\text{SgrA*}}$ $(\mu\mathrm{as})$ & $\displaystyle \Delta_{\rm{HM3}}^{\text{SgrA*}}$ $(\%)$ \\ 
\hline 
\text{Sc.} & $1$ & $5.1962$ & $39.4908$& $5.9744$ & $49.8831$ & $3.7007$ \\ 
\hline 
$1$ & $2$ & $3.8999$ & $29.6397$& $29.4292$ & $37.4397$ & $27.7227$ \\ 
\hline 
$2$ & $2$ & $3.7977$ & $28.8627$ & $31.2793$ & $36.4581$ & $29.6175$ \\ 
\hline 
$3$ & $2$ & $3.6942$ & $28.0756$ & $33.1532$ & $35.4640$ & $31.5367$ \\ 
\hline 
\end{tabular} 
\vskip 0.5cm 
\centering 
\caption{\textit{ADM-normalized critical impact parameters and the corresponding critical angular diameters for HM4, together with their percentage differences from the adopted M87* and Sgr~A* emission ring central values. The fixed-horizon condition $r_{\rm h}=2M$ implies $j=0$ and hence $M_{\rm ADM}=M$ for all three benchmark configurations. All $\alpha>0$ HM4 configurations shown are exploratory and do not belong to the DEC-admissible sector.
}} \label{tab:SGL_metric4} 
\setlength{\tabcolsep}{10pt} 
\begin{tabular}{|c|c|c|c|c|c|c|} 
\hline 
$\alpha$ & $M_{\rm ADM}$ & $\bar{b}_{\rm ph}$ & $\displaystyle d_{\rm{HM4}}^{\text{M87*}}$ $(\mu\mathrm{as})$ & $\displaystyle \Delta_{\rm{HM4}}^{\text{M87*}}$ $(\%)$ & $\displaystyle d_{\rm{HM4}}^{\text{SgrA*}}$ $(\mu\mathrm{as})$ & $\displaystyle \Delta_{\rm{HM4}}^{\text{SgrA*}}$ $(\%)$ \\ 
\hline 
\text{Sc.} & $1$ & $5.1962$ & $39.4908$& $5.9744$ & $49.8831$ & $3.7007$ \\ 
\hline 
$1$ & $1$ & $5.0934$ & $38.7101$ & $7.8332$ & $48.8969$ & $5.6044$ \\ 
\hline 
$2$ & $1$ & $4.9952$ & $37.9641$ & $9.6092$ & $47.9547$ & $7.4233$ \\ 
\hline 
3 & 1 & 4.9016 & 37.2518 & 11.3051 & 47.0550 & 9.1603\\ 
\hline 
\end{tabular} 
\end{table*}

\par Tables~\ref{tab:SGL_metric1}--\ref{tab:SGL_metric4} illustrate the sensitivity of the predicted critical angular diameter to the ADM-normalized critical impact parameter $\bar b_{\rm ph}$. Within the central-value comparison adopted here, the Schwarzschild reference gives percentage differences of approximately $6\%$ and $4\%$ relative to the M87* and Sgr~A* emission ring angular scales, respectively. The effect of the hairy deformation is strongly metric dependent.

\par For HM1, the configurations within the controlled parameter range have smaller values of $\bar b_{\rm ph}$ than Schwarzschild and correspondingly larger central-value differences. The $\alpha=10$ configuration lies outside this domain and is retained only to illustrate the continuation of the geometric trend. For HM2, the configurations with $\alpha=1$ and $\alpha=10$ lie within the adopted controlled comparison range, while $\alpha=20$ is an exploratory extension beyond it. For HM3, all three benchmark configurations satisfy the DEC restriction derived in Sec.~\ref{sec:parameter_space}. The fixed-horizon relation $j=2M/\alpha$ gives $M_{\rm ADM}=2M$, so that ADM normalization substantially reduces the critical impact parameter relative to its value in seed-mass units. The resulting critical angular diameters lie below the Schwarzschild reference throughout the benchmark range, with central-value differences of approximately $30\%$. These differences characterize the HM3 comparison branch and should not be interpreted as observational exclusions. For HM4, the fixed-horizon condition requires $j=0$ and hence $M_{\rm ADM}=M$. The exploratory $\alpha>0$ configurations exhibit a gradual reduction of $\bar b_{\rm ph}$ and of the corresponding critical angular diameter as $\alpha$ increases. Since these configurations lie outside the DEC-admissible sector, their comparison with the EHT angular scales is included only to illustrate the observational imprint of the HM4 geometric deformation.

\par Overall, the comparison demonstrates that ADM normalization is essential when translating the seed-mass geometric results of Sec.~\ref{photon_shadow} into angular scales at fixed observed mass. The quantities reported here provide an illustrative measure of the sensitivity of the predicted shadow scale to the different geometric deformations, but they do not constitute direct EHT constraints on the corresponding hair parameters. Together with the horizon and photon sphere analyzes above, these results show that the deformation can leave distinct signatures in both the causal and null geodesic structure of the spacetime, with the magnitude and direction of the effect depending sensitively on the particular hairy geometry.

\par We now turn to a complementary probe of these spacetimes through their QNM spectra. Whereas the observables considered thus far are determined by the background geometry and its null geodesics, QNMs characterize the response of the spacetime to perturbations. More generally, QNM behavior can be explored using scalar test-fields propagating on curved backgrounds, which provide a useful proxy for qualitative aspects of the underlying perturbative dynamics, without reproducing the full gravitational perturbation spectrum. The following section therefore examines how the same geometric deformations modify the characteristic oscillation frequencies and damping rates for the scalar test-field, and how these trends compare with the photon sphere behavior identified above.


\section{Quasinormal modes of hairy black holes}\label{QNMS}

With the geometric and optical properties of the four hairy black hole metrics discussed above, we now turn to a study of their perturbative behavior via the QNM spectrum of a scalar test-field. We compute the QNM frequencies corresponding to each metric using both the WKB approximation and the AIM. For the AIM analysis, we derive a general form of the radial equation applicable to the class of metric functions considered in this work. We then compare the two approaches by quantifying the percentage difference between the resulting QNM spectra.

Note that throughout this section, we work in natural units, $G_N=c=1$, and fix the mass scale by setting $M=1$. The QNM frequencies are therefore reported in dimensionless units, with the physical scaling restored by $\omega_{\rm phys}\sim \omega/M$. No additional normalization by $M_{\rm ADM}$ is applied in this section. This choice is made so that the QNM calculation can be compared directly across the four metric functions using the same seed-mass normalization adopted in the perturbation equations.


\subsection{The QNM eigenvalue problem}

The theory of QNMs and their applications is extensive; see Refs.~\cite{Nollert:1999ji,Kokkotas:1999bd,refBertiCardoso,refKonoplyaZhidenkoReview} for detailed reviews. Here, we formulate the QNM problem for a massless, minimally coupled scalar test-field propagating on the fixed hairy black hole geometries. This provides a controlled probe of the perturbative properties of the background without requiring the coupled perturbation equations of the gravitational and additional source sectors.

The scalar field $\Phi$ obeys the Klein--Gordon equation
\begin{equation}
\Box\Phi = \frac{1}{\sqrt{-g}} \partial_\mu \left( \sqrt{- g}\,g^{\mu\nu}\partial_\nu\Phi
\right) = 0 \;.
\label{eq:KG_scalar}
\end{equation}
For the static and spherically symmetric geometries considered here, we decompose the scalar test-field into modes,
\begin{equation}
\Phi(t,r,\vartheta,\phi) = \sum^{\infty}_{n=0} \sum^{\infty}_{\ell=0} \sum^{\ell}_{m=-\ell}  e^{-i\omega_{n\ell} t} Y_{\ell m}(\vartheta,\phi) \frac{\psi_{n\ell}(r)}{r} \;,
\label{eq:scalar_decomposition}
\end{equation}
where $Y_{\ell m}$ are the spherical harmonics. The integer $\ell=0,1,2,\ldots$ is the angular multipolar number, while $m=-\ell,\ldots,\ell$ is the corresponding azimuthal number. Owing to spherical symmetry, the QNFs are independent of $m$. For each multipole $\ell$, the QNM spectrum consists of a discrete sequence of complex QNFs labeled by the overtone number $n=0,1,2,\ldots$, with $n=0$ denoting the fundamental mode and $n>0$ the monotonically increasing higher overtones. Accordingly, each QNF is labelled by the pair $(\ell,n)$ and is denoted by $\omega_{\ell n}$. Upon introducing the tortoise coordinate $x$ through
\begin{equation}
\frac{dx}{dr} = \frac{1}{f(r)} \;,
\label{eq:tortoise}
\end{equation}
the radial Klein--Gordon equation becomes
\begin{equation}
\frac{d^2\psi_{\ell n}}{dx^2} + \left[ \omega^2-V_{\ell}(r(x)) \right] \psi_{\ell n} = 0 \;,
\label{mastereq}
\end{equation}
with effective potential
\begin{equation}
V_{\ell}(r) = f(r) \left[ \frac{\ell(\ell+1)}{r^2} + \frac{f'(r)}{r} \right] \;.
\label{potential}
\end{equation}
Here, the prime denotes differentiation with respect to the areal radius $r$. Thus, $V_{\ell}(r)$ denotes the effective potential expressed in terms of $r$, whereas $V_{\ell}(r(x))$ denotes the same function composed with the tortoise-coordinate map.

\par We assume throughout a harmonic time dependence $e^{-i\omega t}$. For the asymptotically flat, non-extremal configurations considered in the QNM analysis, the tortoise coordinate satisfies $x\rightarrow-\infty$ at the event horizon and $x \rightarrow +\infty$ at spatial infinity. QNMs are defined by requiring purely ingoing behavior at the event horizon,
\begin{equation}
\psi_{\ell n} \sim e^{-i\omega x},
\qquad
x\rightarrow-\infty \;,
\label{boundary at horizon}
\end{equation}
and purely outgoing behavior at spatial infinity,
\begin{equation}
\psi_{\ell n} \sim e^{+i\omega x},
\qquad
x\rightarrow+\infty \;.
\label{boundary at infinity}
\end{equation}
These intrinsically dissipative boundary conditions select a discrete set of complex frequencies,
\begin{equation}
\omega_{\ell n} = \omega_{\rm Re}
+
i\omega_{\rm Im} \;,
\label{frequency}
\end{equation}
where recall that $n=0$ denotes the fundamental mode and $n>0$ the successive overtones. With the adopted $e^{-i\omega t}$ convention, damped modes satisfy
\begin{equation}
\omega_{\rm Im}<0 \;.
\end{equation}
The real part $\omega_{\rm Re}$ determines the oscillation frequency, while $-\omega_{\rm Im}$ determines the damping rate. A mode with $\omega_{\rm Im}>0$ would instead grow exponentially in time and signal a linear instability of the scalar test-field.

\par
\begin{figure*}
\centering
\captionsetup[subfigure]{justification=raggedleft,singlelinecheck=false}
\begin{subfigure}[b]{0.475\textwidth}
\centering
\includegraphics[width=\textwidth]{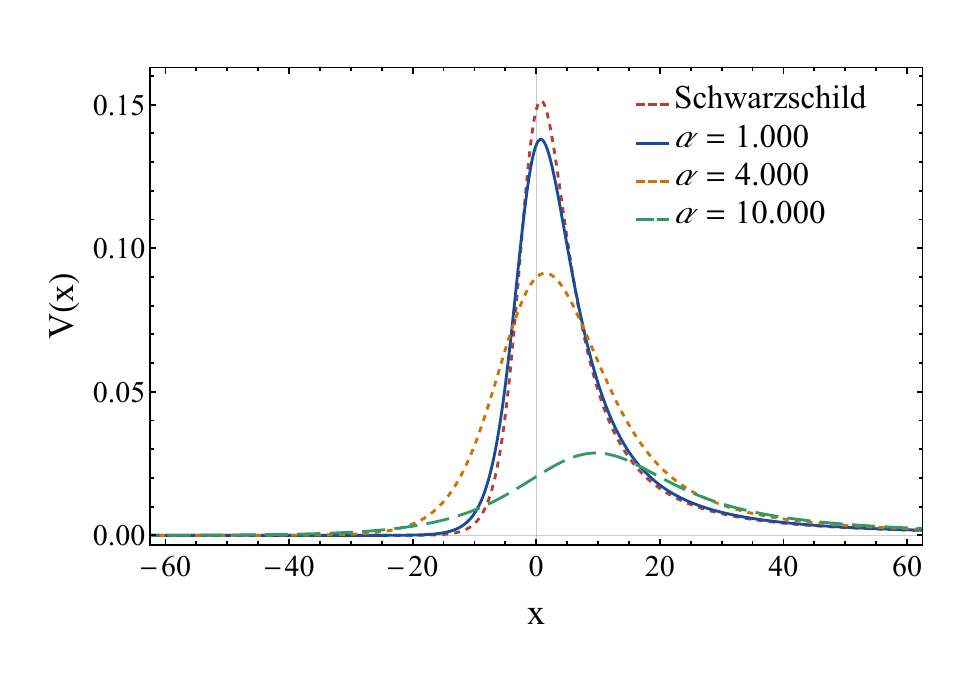}
\vspace{-1cm}
\caption{\it Hairy metric 1}
\label{fig:potential_hm1}
\end{subfigure}
\hfill
\begin{subfigure}[b]{0.475\textwidth}
\centering
\includegraphics[width=\textwidth]{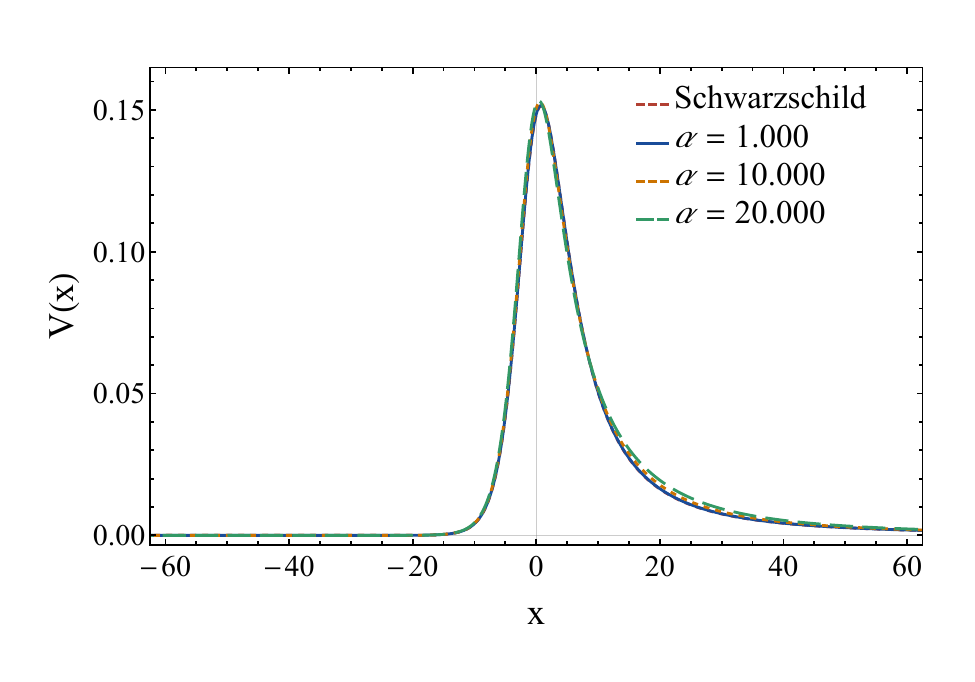}
\vspace{-1cm}
\caption{\it Hairy metric 2}
\label{fig:potential_hm2}
\end{subfigure}
\vspace{0.15cm}
\begin{subfigure}[b]{0.475\textwidth}
\centering
\includegraphics[width=\textwidth]{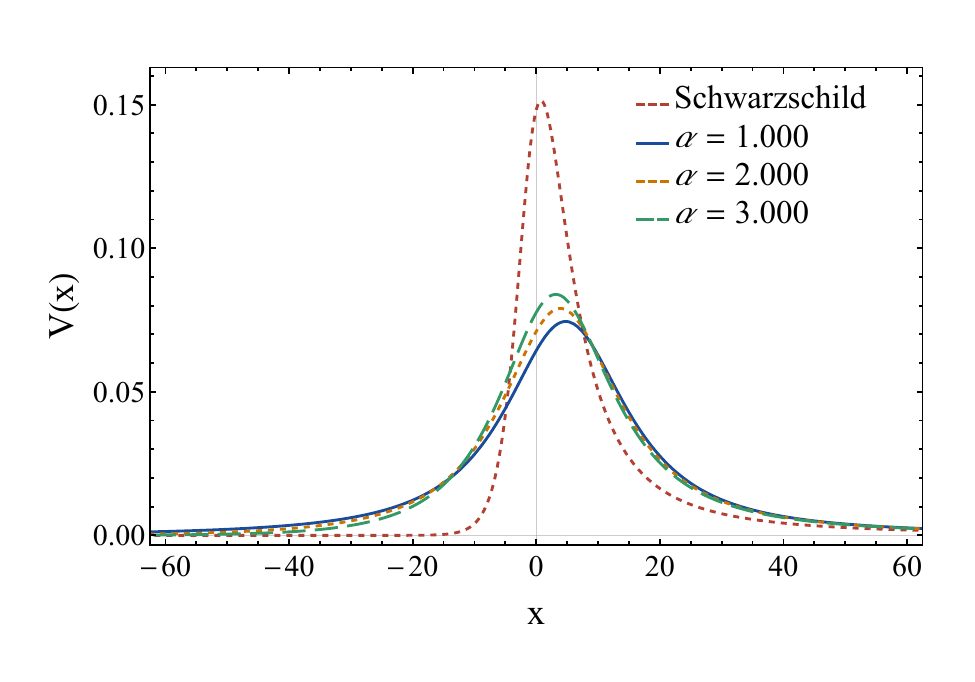}
\vspace{-1cm}
\caption{\it Hairy metric 3}
\label{fig:potential_hm3}
\end{subfigure}
\hfill
\begin{subfigure}[b]{0.475\textwidth}
\centering
\includegraphics[width=\textwidth]{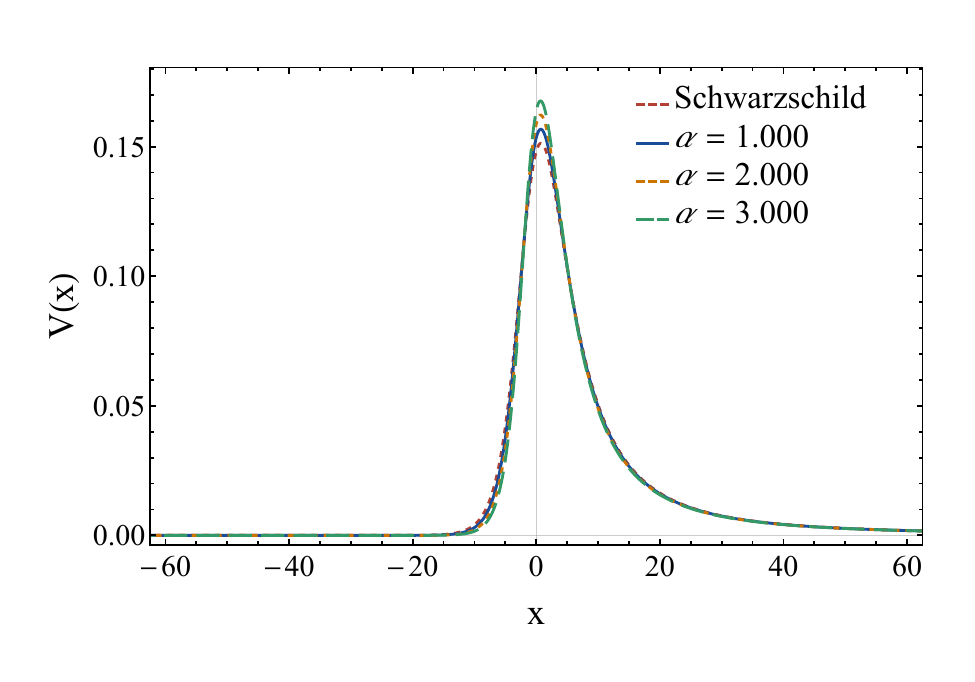}
\vspace{-1cm}
\caption{\it Hairy metric 4}
\label{fig:potential_hm4}
\end{subfigure}
\vskip 0.25cm
\caption{\textit{Effective scalar-field potentials $V_{\ell}(r(x))$ as functions of the tortoise coordinate $x$ for HM1--HM4, with $\ell=2$ and seed mass
$M=1$. The Schwarzschild potential is shown for reference. For HM3, the fixed-horizon condition $r_{\rm h}=2M$ implies $j=2M/\alpha$, while for HM4 it implies $j=0$ for $\alpha>0$. Parameter choices outside the
controlled domains identified in Sec.~\ref{sec:parameter_space}, including $\alpha=10$ for HM1, $\alpha=20$ for HM2, and all nonzero-$\alpha$ HM4 configurations, are shown only for exploratory
comparison.}}
\label{fig:Potential_plots}
\end{figure*}

 Before computing the QNM spectra, it is useful to examine the corresponding effective potentials. Fig.~\ref{fig:Potential_plots} shows $V_{\ell}(r)$ for the four hairy metrics for $\ell=2$ as the deformation parameter $\alpha$ is varied. For the configurations analyzed, the effective potentials retain the smooth single-barrier structure characteristic of a scattering problem. This behavior is particularly favorable for the application of WKB methods, and also provides a useful diagnostic of the radial problem subsequently treated using the AIM. The deformation of the potential height, curvature, and peak location is metric dependent. HM1, HM3, and HM4 exhibit more pronounced changes over the parameter values displayed as $\alpha$ increases, whereas HM2 remains comparatively stable over its controlled range. Within the WKB picture, variations in the height and curvature of the barrier are expected to produce corresponding changes in the oscillation and damping scales of the QNMs, since in WKB-based approximations, the spectrum is determined by the value of the effective potential and its derivatives evaluated at the potential maximum. In particular, changes in the peak height primarily affect the oscillation scale, while changes in the curvature near the maximum influence the damping rate. The spectra calculated below provide the quantitative assessment of these trends.

\subsection{QNM calculations using the WKB method}\label{WKB}

\par The WKB method is one of the standard semi-analytic techniques used for computing black hole QNFs. It is particularly useful for effective potentials with a single peak, since it relates the QNM spectrum directly to the local properties of the potential near its maximum. In this sense, the method provides both a practical approximation scheme and a transparent physical interpretation of wave propagation through the effective potential barrier; since it relates the QNM spectrum to the value and successive derivatives of the effective potential evaluated at its maximum, the WKB method is therefore particularly well suited to the single-peaked effective potential profiles shown in Fig.~\ref{fig:Potential_plots}.

Let us denote the location of the maximum of the effective potential using $x_0$, and define
\begin{equation}
V_0 \equiv V_\ell(x_0) \;, \qquad V_0^{(k)} \equiv \frac{d^k V_\ell}{dx^k} \big \vert_{x=x_0} \;.
\label{eq:wkb_derivatives}
\end{equation}
At the potential maximum,
\begin{equation}
V_0^{(1)}=0 \;, \qquad V_0^{(2)}<0 \;.
\end{equation}
At $N$th WKB order, the QNFs are determined by the quantization condition,
\begin{equation}
\frac{i\left(\omega^2-V_0\right)}{\sqrt{-2V_0^{(2)}}} - \sum_{j=2}^{N}\Lambda_j = n+\frac{1}{2} \;,
\label{eq:wkb_condition}
\end{equation}
where the correction terms $\Lambda_j$ depend on the overtone number $n$ and on higher derivatives of the effective potential evaluated at $x_0$. Since the potentials in Eq.~\eqref{potential} are given explicitly as functions of the areal radius $r$, the required tortoise coordinate derivatives are obtained recursively using Eq.~\eqref{eq:tortoise}. In the numerical calculations presented below, we employ the sixth-order WKB approximation, corresponding to $N=6$ in Eq.~\eqref{eq:wkb_condition}.\footnote{The WKB method was introduced for black hole QNMs at first order in Ref.~\cite{Schutz:1985km} and extended to third order in Ref.~\cite{Iyer:1986np}. The sixth-order extension was developed in Ref.~\cite{Konoplya:2004ip}, while higher-order implementations up to 13th order were obtained in Ref.~\cite{Matyjasek:2017psv}. See Ref.~\cite{Konoplya2019_recipes} for a detailed discussion of the accuracy and limitations of the WKB approach.}

\par The WKB expansion is generally most reliable for low overtones and angular multipoles satisfying $\ell\gg n$, while its accuracy can deteriorate as the overtone number increases, particularly when $n$ becomes comparable to or exceeds $\ell$. Accordingly, the modes considered below with $n<\ell$ lie in the more favorable WKB regime, whereas the $(\ell,n)=(2,2)$ mode lies outside this regime and is retained primarily for comparison with the independent AIM calculation. More generally, the WKB expansion is asymptotic, so increasing the WKB order does not by itself guarantee monotonic convergence to the exact QNF \cite{Konoplya2019_recipes}.

\par Table~\ref{tab:wkb_qnm_models} reports the resulting scalar QNFs for $\ell=2,\ldots,7$ and $n=0,1,2$, using the common seed-mass normalization $M=1$. The four hairy geometries are evaluated at $\alpha=1$ and compared against Schwarzschild and RN reference spacetimes. For the RN benchmark we take $Q/M=0.5$. Note, however, that the RN configuration is included only as a familiar reference geometry; its electromagnetic charge $Q$ should not be identified with the effective charge parameter that appears in the gravitationally-decoupled hairy solutions. For HM3, the fixed-horizon condition $r_{\rm h}=\alpha j=2M$ gives $j=2M/\alpha$, while for HM4 the condition $r_{\rm h}=2M+\alpha j=2M$ gives $j=0$ for $\alpha>0$.

\par The WKB results provide a semianalytic reference against which the AIM frequencies obtained in the following subsection can be compared. The differences between the two calculations are therefore treated as cross-method discrepancies rather than as numerical errors assigned to either method individually. Agreement between the two approaches provides a useful consistency check on the scalar QNM spectra and on their response to the hairy deformations within the adopted $M=1$ normalization. We emphasize, however, that the present calculation concerns scalar test-field perturbations on fixed backgrounds, and therefore does not constitute a full linear stability analysis of the gravitationally-decoupled solutions, which would require perturbing the geometry together with the additional source sector.

\begin{table}[t]
\centering
\caption{\textit{Sixth-order WKB QNFs $M\omega$ of a massless scalar test-field for the Schwarzschild, RN, and hairy black hole geometries, with seed mass $M=1$. The hairy configurations are evaluated at $\alpha=1$, while the RN reference has $Q/M=0.5$. For HM3, the fixed-horizon condition $r_{\rm h}=\alpha j=2M$ implies $j=2M$ at $\alpha=1$, whereas for HM4, $r_{\rm h}=2M+\alpha j=2M$ implies $j=0$. The $\alpha>0$ HM4 configuration is included only as an exploratory geometric benchmark, as discussed in Sec.~\ref{sec:parameter_space}.}}
\label{tab:wkb_qnm_models}
\vskip 0.2cm
\renewcommand{\arraystretch}{1.8}
\setlength{\tabcolsep}{5pt}
\resizebox{\linewidth}{!}{%
\begin{tabular}{|c|c|c|c|c|c|c|c|c|c|c|c|c|c|}
\hline
\multirow{2}{*}{$\ell$} & \multirow{2}{*}{$n$}
& \multicolumn{2}{c|}{\text{Schwarzschild}}
& \multicolumn{2}{c|}{\text{RN ($Q=0.5$)}}
& \multicolumn{2}{c|}{\text{HM1}}
& \multicolumn{2}{c|}{\text{HM2}}
& \multicolumn{2}{c|}{\text{HM3}}
& \multicolumn{2}{c|}{\text{HM4}} \\
\cline{3-14}
& 
& \text{Re $\omega$} & \text{Im $\omega$}
& \text{Re $\omega$} & \text{Im $\omega$}
& \text{Re $\omega$} & \text{Im $\omega$}
& \text{Re $\omega$} & \text{Im $\omega$}
& \text{Re $\omega$} & \text{Im $\omega$}
& \text{Re $\omega$} & \text{Im $\omega$} \\
\hline
\multirow{3}{*}{2}
& 0 & 0.4836 & $-0.0968$
      & 0.5059 & $-0.0979$
      & 0.4539 & $-0.0824$
      & 0.4820 & $-0.0966$
      & 0.3213  & $-0.0449$
      & 0.4928  & $-0.1018$ \\
& 1 & 0.4639 & $-0.2956$
      & 0.4873 & $ -0.2988$
      & 0.4421 & $-0.2504$
      & 0.4609 & $-0.2950$
      & 0.3124  & $-0.1362$
      & 0.4691  & $-0.3114$ \\
& 2 & 0.4304 & $-0.5087$
      & 0.4558 & $-0.5129$
      & 0.4249 & $-0.4230$
      & 0.4236 & $-0.5078$
      & 0.2954  & $-0.2317$
      & 0.4272  & $-0.5375$ \\
\hline
\multirow{3}{*}{3}
& 0 & 0.6754 & $-0.0965$
      & 0.7065 & $-0.0977$
      & 0.6336 & $-0.0822$
      & 0.6734 & $-0.0964$
      & 0.4493  & $-0.0449$
      & 0.6885  & $-0.1016$ \\
& 1 & 0.6607 & $-0.2923$
      & 0.6926  & $-0.2957$
      & 0.6243 & $-0.2482$
      & 0.6579 & $-0.2919$
      & 0.4428  & $-0.1353$
      & 0.6710  & $-0.3079$ \\
& 2 & 0.6336 & $-0.4960$
      & 0.6671 & $-0.5011$
      & 0.6079 & $-0.4177$
      & 0.6286 & $-0.4954$
      & 0.4302  & $-0.2278$
      & 0.6383  & $-0.5232$ \\
\hline
\multirow{3}{*}{4}
& 0 & 0.8674 & $-0.0964$
      & 0.9073 & $-0.0976$
      & 0.8136 & $-0.0821$
      & 0.8651 & $-0.0963$
      & 0.5773  & $-0.0448$
      & 0.8845  & $-0.1015$ \\
& 1 & 0.8558 & $-0.2909$
      & 0.8964 & $-0.2944$
      & 0.8061 & $-0.2473$
      & 0.8528 & $-0.2907$
      & 0.5723  & $-0.1349$
      & 0.8708  & $-0.3064$ \\
& 2 & 0.8337 & $-0.4903$
      & 0.8755 & $-0.4958$
      & 0.7921 & $-0.4149$
      & 0.8292 & $-0.4899$
      & 0.5624  & $-0.2263$
      & 0.8443  & $-0.5169$ \\
\hline
\multirow{3}{*}{5}
& 0 & 1.0596 & $-0.0963$
      & 1.1083 & $-0.0975$
      & 0.9937 & $-0.0821$
      & 1.0569 & $-0.0963$
      & 0.7055  & $-0.0448$
      & 1.0807  & $-0.1015$ \\
& 1 & 1.0500 & $-0.2902$
      & 1.0993 & $-0.2937$
      & 0.9875 & $-0.2469$
      & 1.0468 & $-0.2899$
      & 0.7014  & $-0.1347$
      & 1.0694  & $-0.3056$ \\
& 2 & 1.0315 & $-0.4873$
      & 1.0818 & $-0.4930$
      & 0.9756 & $-0.4134$
      & 1.0271 & $-0.4870$
      & 0.6932  & $-0.2255$
      & 1.0473  & $-0.5136$ \\
\hline
\multirow{3}{*}{6}
& 0 & 1.2518 & $-0.0963$
      & 1.3094 & $-0.0975$
      & 1.1739 & $-0.0821$
      & 1.2488 & $-0.0963$
      & 0.8336  & $-0.0448$
      & 1.2769  & $-0.1014$ \\
& 1 & 1.2438 & $-0.2897$
      & 1.3018 & $-0.2933$
      & 1.1686 & $-0.2467$
      & 1.2402 & $-0.2896$
      & 0.8301  & $-0.1346$
      & 1.2673  & $-0.3052$ \\
& 2 & 1.2278 & $-0.4856$
      & 1.2868 & $-0.4914$
      & 1.1583 & $-0.4125$
      & 1.2233 & $-0.4854$
      & 0.8232  & $-0.2251$
      & 1.2484  & $-0.5117$ \\
\hline
\multirow{3}{*}{7}
& 0 & 1.4442 & $-0.0963$
      & 1.5106 & $-0.0975$
      & 1.3543 & $-0.0821$
      & 1.4407 & $-0.0963$
      & 0.9618  & $-0.0448$
      & 1.4731  & $-0.1014$ \\
& 1 & 1.4371 & $-0.2895$
      & 1.5039 & $-0.2931$
      & 1.3496 & $-0.2465$
      & 1.4332 & $-0.2894$
      & 0.9588  & $-0.1346$
      & 1.4648  & $-0.3049$ \\
& 2 & 1.4232 & $-0.4845$
      & 1.4908 & $-0.4904$
      & 1.3405 & $-0.4119$
      & 1.4185 & $-0.4843$
      & 0.9527  & $-0.2248$
      & 1.4483 & $-0.5105$ \\
\hline
\end{tabular}}
\end{table}


\subsection{QNM calculations from the Asymptotic Iteration Method}\label{AIM}

\par Another method for calculating QNMs is a semi-analytic approach based on the iterative structure of second-order linear differential equations, known as the AIM \cite{Cho:2009cj,Cho:2011sf}. The method proceeds by recasting the perturbation equation into a second-order form whose coefficient functions can be iterated to obtain a quantization condition for the eigenvalue $\omega$. Since the objective here is to apply the AIM uniformly to the hairy black hole metrics introduced above, we first summarize the essential steps before deriving the specific coefficient functions used in our calculation.

The standard AIM form is written as
\begin{equation}
\label{AIM SODE}
\chi''(\xi) = \lambda_0(\xi)\chi'(\xi) + s_0(\xi)\chi(\xi) \;,
\end{equation}
where $\lambda_0(\xi)$ and $s_0(\xi)$ are smooth coefficient functions and primes denote differentiation with respect to the radial variable $\xi$. Repeated differentiation preserves this structure, giving
\begin{equation}
\chi^{(k+2)}(\xi) = \lambda_k(\xi)\chi'(\xi) + s_k(\xi)\chi(\xi) \;,
\end{equation}
with the recursion relations
\begin{align}
\lambda_k(\xi) &= \lambda'_{k-1}(\xi) + s_{k-1}(\xi) + \lambda_0(\xi)\lambda_{k-1}(\xi) \;,
\\
s_k(\xi) &= s'_{k-1}(\xi) + s_0(\xi)\lambda_{k-1}(\xi) \;.
\end{align}
Here, $k=1,2,\ldots$ denotes the AIM iteration index. The corresponding quantization condition is
\begin{equation}
\delta_k(\xi) \equiv s_k(\xi)\lambda_{k-1}(\xi) - s_{k-1}(\xi)\lambda_k(\xi) = 0 \;,
\label{eq:AIM_quantisation}
\end{equation}
which determines the complex QNFs at sufficiently large iteration order.

\par In the improved AIM, repeated numerical differentiation is avoided by expanding the coefficient functions about a fixed point $\xi_0$ \cite{Cho:2009cj,Cho:2011sf},
\begin{align}
\lambda_k(\xi) &= \sum_{i=0}^{\infty} c_k^{\,i}(\xi-\xi_0)^i \;,
\\
s_k(\xi) &= \sum_{i=0}^{\infty} d_k^{\,i}(\xi-\xi_0)^i \;,
\end{align}
where $c_k^{\,i}$ and $d_k^{\,i}$ are the corresponding Taylor coefficients. Substitution into the AIM recursion relations gives
\begin{align}
c_k^{\,i} &= (i+1)c_{k-1}^{\,i+1} + d_{k-1}^{\,i} + \sum_{j=0}^{i} c_0^{\,j}c_{k-1}^{\,i-j} \;,
\\
d_k^{\,i} &= (i+1)d_{k-1}^{\,i+1} + \sum_{j=0}^{i} d_0^{\,j}c_{k-1}^{\,i-j} \;.
\end{align}
The quantization condition then reduces to
\begin{equation}
d_k^{\,0}c_{k-1}^{\,0} - d_{k-1}^{\,0}c_k^{\,0} = 0 \;.
\label{eq:improved_AIM_quantisation}
\end{equation}

\par We now adapt this construction to the asymptotically flat, non-extremal black hole geometries considered in this work. Rather than derive the AIM equations separately for each background, we formulate the procedure directly in terms of a generic static and spherically symmetric metric function $f(r)$. Let us begin by introducing the compact coordinate,
\begin{equation} \label{eq:AIM_compact_coordinate}
\xi = 1 - \frac{r_{\rm h}}{r} \;, 
\end{equation}
such that $\xi = 0$ maps $r$ to the event horizon and $\xi \rightarrow 1$ corresponds to spatial infinity. Using the tortoise coordinate Eq.~\eqref{eq:tortoise}, we write 
\begin{equation}
\frac{d}{dx} = f(r)\frac{d}{dr} = A(\xi)\frac{d}{d\xi} \;, \qquad
A(\xi) \equiv \frac{(1-\xi)^2}{r_{\rm h}}f(r(\xi)) \;.
\end{equation}
\noindent The radial equation, Eq.~\eqref{mastereq}, can then be written as
\begin{equation}
\varphi''\left(\xi \right) + \frac{A'(\xi)}{A(\xi)}\varphi'\left(\xi \right) + \frac{\omega^2-V(r(\xi))}{A^2(\xi)}\varphi\left(\xi \right)=0 \;,
\end{equation}
where the change of radial coordinate is made explicit through $\varphi(\xi)\equiv\psi(r(\xi)).$

\par We factor out the QNM boundary behavior according to
\begin{equation} \label{eq:varphixi}
\varphi(\xi)=P(\xi)\chi(\xi) \;,
\end{equation}
where for a non-extremal horizon,
\begin{equation}
f(r)\simeq f'(r_{\rm h})(r-r_{\rm h}) \;.
\end{equation}
The ingoing boundary condition $\varphi\left(\xi\right)\sim e^{-i\omega x}$ then gives
\begin{equation}
\varphi\left(\xi\right) \sim \xi^{-i\omega/f'(r_{\rm h})} \;, \qquad \xi \to 0 \;;
\end{equation}
\noindent at spatial infinity, the outgoing boundary condition becomes
\begin{equation}
\varphi\left(\xi\right) \sim
\exp \left[ \frac{i\omega r_{\rm h}}{1-\xi}\right] (1-\xi)^{-2 i \omega M_{\rm ADM}} \;, \qquad \xi \rightarrow 1 \;.
\end{equation}
Note that this expression is obtained under the assumption that
\begin{equation}
f(r)=1-\frac{2M_{\rm ADM}}{r}+\mathcal{O}(r^{-2}) \;,
\end{equation}
so that the tortoise coodinate can be expressed as
\begin{equation}
x=r+2M_{\rm ADM}\ln r+\mathcal{O}(r^{-1}) \;.
\end{equation}

\par To capture this ingoing and outgoing behavior via Eq.~\eqref{eq:varphixi}, we use
\begin{equation}
P(\xi)=\xi^{-i\omega/f'(r_{\rm h})}(1-\xi)^{-2i\omega M_{\rm ADM}} \exp\left[\frac{i\omega r_{\rm h}}{1-\xi}\right] \;.
\end{equation}
Upon introducing
\begin{equation}
g(\xi)\equiv\frac{P'(\xi)}{P(\xi)} = \frac{a}{\xi}-\frac{b}{1-\xi}+\frac{c}{(1-\xi)^2} \;,
\end{equation}
for which
\begin{equation}
a=-\frac{i\omega}{f'(r_{\rm h})} \;, \qquad b=-2i\omega M_{\rm ADM} \;, \qquad c=i\omega r_{\rm h} \;,
\end{equation}
the equation for $\chi$ can be written in the standard AIM form,
\begin{equation}
\chi''=\lambda_0 (\xi) \chi' + s_0 (\xi) \chi \;.
\end{equation}
Here, however,
\begin{equation}
\label{eq:generalAIM}
\begin{aligned}
\lambda_0(\xi) & = -\left[2g(\xi)+\frac{A'(\xi)}{A(\xi)}\right] \;,
\\
s_0(\xi) & = -\left[g'(\xi)+g^2(\xi)+\frac{A'(\xi)}{A(\xi)}g(\xi)+\frac{\omega^2-V(r(\xi))}{A^2(\xi)}\right] \;.
\end{aligned}
\end{equation}
\noindent Note that primes of $\chi$ denote derivatives with respect to $\xi$. 

\par This serves as a common AIM formulation for all the non-extremal, asymptotically flat geometries considered in this work. The QNM boundary conditions are contained entirely in $P(\xi)$, while the remaining function $\chi(\xi)$ is determined using the improved AIM recursion on the compact interval $0 \leq \xi < 1$. As specified at the beginning of this subsection, for the improved AIM, the recursion relations are evaluated by expanding about a fixed point $\xi_0$ in the exterior region. For each metric and multipole, we choose this point to correspond to the maximum of the effective potential. Specifically, if $r_0>r_{\rm h}$ satisfies
\begin{equation}
V'(r_0) = 0 \;, \qquad V ''(r_0) < 0 \;,
\end{equation}
then we take
\begin{equation}
\xi_0 = 1-\frac{r_{\rm h}}{r_0} \;.
\label{eq:AIM_expansion_point}
\end{equation}
This choice provides a simple and uniform prescription for selecting the AIM expansion point across the different black hole geometries considered in this work.

\par In the numerical implementation, the Taylor expansions of $\lambda_0(\xi)$ and $s_0(\xi)$ are constructed about $\xi_0$. We use $N=75$ AIM iterations and truncate the Taylor coefficient recursion at order $K=85$, with 150-digit working precision. The complex QNFs are obtained from the simultaneous conditions
\begin{equation}
{\rm Re}\,\delta_N(\omega)=0 \;,
\qquad
{\rm Im}\,\delta_N(\omega)=0 \;.
\end{equation}
For each targeted pair $(\ell,n)$, the complex root search is initialized using an estimate based on the corresponding WKB QNF.

\par In Table~\ref{tab:aim_qnm_models}, we present the QNM spectra obtained using the improved AIM together with the general expressions for $\lambda_0$ and $s_0$ derived above. We observe close agreement between the AIM and WKB results for the combinations of $\ell$ and $n$ considered here. This agreement is quantified in Table~\ref{tab:WKB_AIM_PERCENT_ERROR} through the relative percentage discrepancies
\begin{align}
\Delta_{\rm Re} &= 100 \frac{
\left|
{\rm Re}\,\omega_{\rm AIM}
-
{\rm Re}\,\omega_{\rm WKB}
\right|
}{\left|
{\rm Re}\,\omega_{\rm WKB}
\right|
} \;,
\\
\Delta_{\rm Im}
&= 100
\frac{
\left|
{\rm Im}\,\omega_{\rm AIM}
-
{\rm Im}\,\omega_{\rm WKB}
\right|
}{
\left|
{\rm Im}\,\omega_{\rm WKB}
\right|
} \;.
\end{align}
\par At the precision reported in Table~\ref{tab:WKB_AIM_PERCENT_ERROR}, the WKB and improved AIM calculations show close agreement across the modes considered. The largest tabulated discrepancy is approximately $0.33\%$, occurring in the imaginary part of the HM1 $(\ell,n)=(2,2)$ mode, for which the WKB approximation is expected to be comparatively less accurate. The agreement is generally stronger for the remaining modes, providing a useful cross-method consistency check on the scalar QNM trends reported here.

\par Taken together, the WKB and improved AIM calculations provide consistent determinations of the scalar test-field QNM spectra within the seed-mass normalization adopted here. Their agreement supports the metric-dependent trends identified above, while the present test-field analysis and the finite set of modes considered do not constitute a complete stability analysis of the underlying gravitationally-decoupled solutions. In the next section, we discuss how these QNF results (together with the complementary horizon, photon sphere, and shadow-scale diagnostics considered in the preceding sections) provide a unified characterization of the effects of the different geometric deformations across the four hairy black hole families.

\begin{table}[t]
\centering
\caption{\textit{Improved AIM scalar QNFs $M\omega$ for the Schwarzschild, RN, and hairy black hole geometries, with seed mass $M=1$. The hairy configurations are evaluated at $\alpha=1$, while the RN reference has $Q/M=0.5$. For HM3, the fixed-horizon condition $r_{\rm h}=\alpha j=2M$ implies $j=2M$ at $\alpha=1$, whereas for HM4, $r_{\rm h}=2M+\alpha j=2M$ implies $j=0$. The $\alpha>0$ HM4 configuration is included only as an exploratory geometric benchmark, as discussed in Sec.~\ref{sec:parameter_space}.}}
\label{tab:aim_qnm_models}
\vskip 0.2cm
\renewcommand{\arraystretch}{1.8}
\setlength{\tabcolsep}{5pt}
\resizebox{\linewidth}{!}{%
\begin{tabular}{|c|c|c|c|c|c|c|c|c|c|c|c|c|c|}
\hline
\multirow{2}{*}{\text{$\ell$}} & \multirow{2}{*}{\text{$n$}}
& \multicolumn{2}{c|}{\text{Schwarzschild}}
& \multicolumn{2}{c|}{\text{RN ($Q=0.5$)}}
& \multicolumn{2}{c|}{\text{HM1}}
& \multicolumn{2}{c|}{\text{HM2}}
& \multicolumn{2}{c|}{\text{HM3}}
& \multicolumn{2}{c|}{\text{HM4}} \\
\cline{3-14}
& 
& \text{Re $\omega$} & \text{Im $\omega$}
& \text{Re $\omega$} & \text{Im $\omega$}
& \text{Re $\omega$} & \text{Im $\omega$}
& \text{Re $\omega$} & \text{Im $\omega$}
& \text{Re $\omega$} & \text{Im $\omega$}
& \text{Re $\omega$} & \text{Im $\omega$} \\
\hline
\multirow{3}{*}{2}
& 0 & 0.4836 & $-0.0968$
      & 0.5059 & $-0.0979 $
      & 0.4539 & $-0.0824$
      & 0.4821 & $-0.0966$
      & 0.3213  & $-0.0449$
      & 0.4928  & $-0.1018$ \\
& 1 & 0.4639 & $-0.2956$
      & 0.4873 & $ -0.2988$
      & 0.4419 & $-0.2502$
      & 0.4609 & $-0.2950$
      & 0.3124  & $-0.1362$
      & 0.4691  & $-0.3114$ \\
& 2 & 0.4305 & $-0.5086$
      & 0.4558 & $-0.5129$
      & 0.4242 & $-0.4244$
      & 0.4237 & $-0.5076$
      & 0.2954  & $-0.2317$
      & 0.4274  & $-0.5374$ \\
\hline
\multirow{3}{*}{3}
& 0 & 0.6754 & $-0.0965$
      & 0.7065 & $-0.0977$
      & 0.6336 & $-0.0822$
      & 0.6734 & $-0.0964$
      & 0.4493  & $-0.0449$
      & 0.6885  & $-0.1016$ \\
& 1 & 0.6607 & $-0.2923$
      & 0.6926  & $-0.2957$
      & 0.6242 & $-0.2481$
      & 0.6579 & $-0.2919$
      & 0.4428  & $-0.1353$
      & 0.6711  & $-0.3079$ \\
& 2 & 0.6336 & $-0.4960$
      & 0.6671 & $-0.5011$
      & 0.6079 & $-0.4179$
      & 0.6286 & $-0.4954$
      & 0.4302  & $-0.2278$
      & 0.6383  & $-0.5232$ \\
\hline
\multirow{3}{*}{4}
& 0 & 0.8674 & $-0.0964$
      & 0.9073 & $-0.0976 $
      & 0.8136 & $-0.0821$
      & 0.8651 & $-0.0963$
      & 0.5773  & $-0.0448$
      & 0.8845  & $-0.1015$ \\
& 1 & 0.8558 & $-0.2909$
      & 0.8964 & $-0.2944$
      & 0.8061 & $-0.2473$
      & 0.8528 & $-0.2907$
      & 0.5723  & $-0.1349$
      & 0.8708  & $-0.3064$ \\
& 2 & 0.8337 & $-0.4903$
      & 0.8755 & $-0.4958$
      & 0.7921 & $-0.4150$
      & 0.8292 & $-0.4899$
      & 0.5624  & $-0.2263$
      & 0.8443  & $-0.5169$ \\
\hline
\multirow{3}{*}{5}
& 0 & 1.0596 & $-0.0963$
      & 1.1083 & $-0.0975$
      & 0.9937 & $-0.0821$
      & 1.0569 & $-0.0962$
      & 0.7055  & $-0.0448$
      & 1.0807  & $-0.1015$ \\
& 1 & 1.0500 & $-0.2902$
      & 1.0993 & $-0.2937 $
      & 0.9875 & $-0.2469$
      & 1.0468 & $-0.2899$
      & 0.7013  & $-0.1347$
      & 1.0694  & $-0.3056$ \\
& 2 & 1.0315 & $-0.4873$
      & 1.0818 & $-0.4930 $
      & 0.9756 & $-0.4135$
      & 1.0271 & $-0.4870$
      & 0.6932  & $-0.2255$
      & 1.0473  & $-0.5136$ \\
\hline
\multirow{3}{*}{6}
& 0 & 1.2519 & $-0.0963$
      & 1.3094 & $-0.0975$
      & 1.1739 & $-0.0821$
      & 1.2488 & $-0.0963$
      & 0.8336 & $-0.0448$
      & 1.2769 & $-0.1014$ \\
& 1 & 1.2438 & $-0.2897$
      & 1.3018 & $-0.2933$
      & 1.1686 & $-0.2467$
      & 1.2402 & $-0.2896$
      & 0.8301  & $-0.1346$
      & 1.2673  & $-0.3052$ \\
& 2 & 1.2278 & $-0.4856$
      & 1.2868 & $-0.4914 $
      & 1.1583 & $-0.4126$
      & 1.2233 & $-0.4854$
      & 0.8232 & $-0.2251$
      & 1.2484  & $-0.5117$ \\
\hline
\multirow{3}{*}{7}
& 0 & 1.4442 & $-0.0963$
      & 1.5106 & $-0.0975 $
      & 1.3543 & $-0.0821$
      & 1.4407 & $-0.0963$
      & 0.9618  & $-0.0449$
      & 1.4731  & $-0.1014$ \\
& 1 & 1.4371 & $-0.2895$
      & 1.5039 & $-0.2931$
      & 1.3496 & $-0.2465$
      & 1.4332 & $-0.2894$
      & 0.9589  & $-0.1347$
      & 1.4648  & $-0.3049$ \\
& 2 & 1.4232 & $-0.4845$
      & 1.4908 & $-0.4904$
      & 1.3405 & $-0.4119$
      & 1.4185 & $-0.4843$
      & 0.9530  & $-0.2248$
      & 1.4483  & $-0.5105$ \\    
\hline
\end{tabular}}
\end{table}

\begin{table}[t]
\centering
\caption{\textit{Relative percentage discrepancies between the WKB and improved AIM scalar QNFs reported in Tables~\ref{tab:wkb_qnm_models} and~\ref{tab:aim_qnm_models}, respectively. The real and imaginary parts are evaluated separately using the WKB result as the reference denominator. Entries reported as $0.00\%$ indicate agreement at the numerical precision used to construct the table and should not be interpreted as exact equality.}}
\label{tab:WKB_AIM_PERCENT_ERROR}
\vskip 0.2cm
\renewcommand{\arraystretch}{1.5}
\setlength{\tabcolsep}{7.5pt}
\resizebox{\textwidth}{!}{%
\begin{tabular}{|c|c|c|c|c|c|c|c|c|c|c|c|c|c|}
\hline
\multirow{2}{*}{\text{$\ell$}} & \multirow{2}{*}{\text{$n$}}
& \multicolumn{2}{c|}{\text{Schwarzschild}}
& \multicolumn{2}{c|}{\text{RN ($Q=0.5$)}}
& \multicolumn{2}{c|}{\text{HM1}}
& \multicolumn{2}{c|}{\text{HM2}}
& \multicolumn{2}{c|}{\text{HM3}}
& \multicolumn{2}{c|}{\text{HM4}} \\
\cline{3-14}
& 
& \text{Re $\omega$} & \text{Im $\omega$}
& \text{Re $\omega$} & \text{Im $\omega$}
& \text{Re $\omega$} & \text{Im $\omega$}
& \text{Re $\omega$} & \text{Im $\omega$}
& \text{Re $\omega$} & \text{Im $\omega$}
& \text{Re $\omega$} & \text{Im $\omega$} \\
\hline
\multirow{3}{*}{2}
& 0 & 0.00 & $0.00$
      & 0.00 & $0.00$
      & 0.00 & $0.00$
      & 0.02 & $0.00$
      & 0.00 & $0.00$
      & 0.00 & $0.00$ \\
& 1 & 0.00 & $0.00$
      & 0.00 & $0.00$
      & 0.05 & $0.08$
      & 0.00 & $0.00$
      & 0.00 & $0.00$
      & 0.00 & $0.00$ \\
& 2 & 0.02 & $0.02$
      & 0.00 & $0.00$
      & 0.16 & $0.33$
      & 0.02 & $0.04$
      & 0.00 & $0.00$
      & 0.05 & $0.02$ \\
\hline
\multirow{3}{*}{3}
& 0 & 0.00 & $0.00$
      & 0.00 & $0.00$
      & 0.00 & $0.00$
      & 0.00 & $0.00$
      & 0.00 & $0.00$
      & 0.00 & $0.00$ \\
& 1 & 0.00 & $0.00$
      & 0.00 & $0.00$
      & 0.02 & $0.04$
      & 0.00 & $0.00$
      & 0.00 & $0.00$
      & 0.01 & $0.00$ \\
& 2 & 0.00 & $0.00$
      & 0.00 & $0.00$
      & 0.00 & $0.05$
      & 0.00 & $0.00$
      & 0.00 & $0.00$
      & 0.00 & $0.00$ \\
\hline
\multirow{3}{*}{4}
& 0 & 0.00 & $0.00$
      & 0.00 & $0.00$
      & 0.00 & $0.00$
      & 0.00 & $0.00$
      & 0.00 & $0.00$
      & 0.00 & $0.00$ \\
& 1 & 0.00 & $0.00$
      & 0.00 & $0.00$
      & 0.00 & $0.00$
      & 0.00 & $0.00$
      & 0.00 & $0.00$
      & 0.00 & $0.00$ \\
& 2 & 0.00 & $0.00$
      & 0.00 & $0.00$
      & 0.00 & $0.02$
      & 0.00 & $0.00$
      & 0.00 & $0.00$
      & 0.00 & $0.00$ \\
\hline
\multirow{3}{*}{5}
& 0 & 0.00 & $0.00$
      & 0.00 & $0.00$
      & 0.00 & $0.00$
      & 0.00 & $0.10$
      & 0.00 & $0.00$
      & 0.00 & $0.00$ \\
& 1 & 0.00 & $0.00$
      & 0.00 & $0.00$
      & 0.00 & $0.00$
      & 0.00 & $0.00$
      & 0.01 & $0.00$
      & 0.00 & $0.00$ \\
& 2 & 0.00 & $0.00$
      & 0.00 & $0.00$
      & 0.00 & $0.02$
      & 0.00 & $0.00$
      & 0.00 & $0.00$
      & 0.00 & $0.00$ \\
\hline
\multirow{3}{*}{6}
& 0 & 0.01 & $0.00$
      & 0.00 & $0.00$
      & 0.00 & $0.00$
      & 0.00 & $0.00$
      & 0.00 & $0.00$
      & 0.00 & $0.00$ \\
& 1 & 0.00 & $0.00$
      & 0.00 & $0.00$
      & 0.00 & $0.00$
      & 0.00 & $0.00$
      & 0.00 & $0.00$
      & 0.00 & $0.00$ \\
& 2 & 0.00 & $0.00$
      & 0.00 & $0.00$
      & 0.00 & $0.02$
      & 0.00 & $0.00$
      & 0.00 & $0.00$
      & 0.00 & $0.00$ \\
\hline
\multirow{3}{*}{7}
& 0 & 0.00 & $0.00$
      & 0.00 & $0.00$
      & 0.00 & $0.00$
      & 0.00 & $0.00$
      & 0.00 & $0.22$
      & 0.00 & $0.00$ \\
& 1 & 0.00 & $0.00$
      & 0.00 & $0.00$
      & 0.00 & $0.00$
      & 0.00 & $0.00$
      & 0.01 & $0.07$
      & 0.00 & $0.00$ \\
& 2 & 0.00 & $0.00$
      & 0.00 & $0.00$
      & 0.00 & $0.00$
      & 0.00 & $0.00$
      & 0.03 & $0.00$
      & 0.00 & $0.00$ \\
\hline
\end{tabular}}
\end{table}

\section{Discussion}\label{Discussion}

\par Any phenomenological analysis of gravitationally-decoupled black hole metrics is meaningful only insofar as the underlying geometries occupy physically admissible regions of parameter space. In this work, we examined physically motivated ranges for the deformation parameters $\alpha$ and $j$ using the relevant energy conditions, the horizon structure, and, where applicable, the RN-like effective interpretation inherited from the parent family specified in Eq.~\eqref{eq:fDEC}. Although alternative notions of admissibility may be considered, these criteria provide a consistent basis for distinguishing controlled configurations from the exploratory benchmarks retained for comparison.

\par As established in Sec.~\ref{sec:parameter_space}, the four geometries studied in this work differ in the status of their admissible parameter domains, reflecting their distinct constructions (reviewed in Sec.~\ref{GD}): HM1 arises from the SEC branch, whereas HM2--HM4 belong to the black hole family constructed under the DEC. HM1 admits the controlled fixed-horizon range $0\leq\alpha<e^2$, beyond which $r_{\rm h}=2M$ ceases to be the outer event horizon. HM2 retains $r_{\rm h}=2M$ throughout its analytic branch, with the adopted range $0\leq\alpha<2e^2$ arising primarily from preservation of its RN-like interpretation rather than from an independent horizon or DEC bound. For HM3, the fixed-horizon condition correlates the two deformation parameters through $j=2M/\alpha$, while the DEC requirements yield $0<\alpha\leq2e^2$; the benchmark values $\alpha=1,2,3$ therefore all lie within the admissible domain. HM4 is qualitatively distinct from the other three families. Although the fixed-horizon condition selects $j=0$ for $\alpha>0$, no nontrivial HM4 configuration in this analytic subclass is DEC-admissible, and relaxing the horizon condition does not alter this conclusion. Accordingly, the controlled physical analysis is provided by HM1--HM3, whereas the $\alpha>0$ HM4 results are retained only as exploratory geometric and perturbative benchmarks. The parameter restrictions therefore have different origins across the four families and should be taken into account when comparing the trends observed within our results.

\par Our first study towards the phenomenological features of these metrics concern their null geodesic properties, summarized in Table~\ref{tab:photon_rings_and_BH_Shadows}. In the common seed-mass normalization $M=1$, HM1 displays an outward displacement of the photon sphere together with an increasing critical impact parameter. The controlled HM1 configurations therefore already differ appreciably from Schwarzschild, while the $\alpha=10$ point illustrates the more pronounced change that accompanies departure from the fixed-horizon branch. HM2 shows a different response: $r_{\rm ph}$ decreases mildly as $\alpha$ increases, whereas $b_{\rm ph}$ increases from $5.2083$ at $\alpha=1$ to $5.3154$ at $\alpha=10$ and $5.4294$ at the exploratory value of $\alpha=20$. For HM3, both $r_{\rm ph}$ and $b_{\rm ph}$ decrease across the DEC-admissible benchmark sequence, although their seed-normalized values remain larger than the Schwarzschild values. The exploratory HM4 sequence instead exhibits modest monotonic reductions of both quantities. In every configuration listed in Table~\ref{tab:photon_rings_and_BH_Shadows}, the relevant photon sphere remains outside the event horizon.  

\par Note, however, that the seed-mass comparison alone is insufficient when these geometric quantities are translated into angular scales for astrophysical black holes with a fixed mass inferred from observations. As shown explicitly in Table~\ref{tab:adm_masses}, the ADM mass differs from $M$ according to
\begin{equation}
M_{\rm ADM}^{\rm HM1}
=
M\left(1+\frac{\alpha}{e^2}\right) \;,
\qquad
M_{\rm ADM}^{\rm HM2}
=
M\left(1+\frac{\alpha}{2e^2}\right) \;,
\end{equation}
while
\begin{equation}
M_{\rm ADM}^{\rm HM3}
=
M_{\rm ADM}^{\rm HM4}
=
M+\frac{\alpha j}{2} \;.
\end{equation}
This distinction is particularly important for HM3, for which the fixed-horizon relation $j=2M/\alpha$ gives $M_{\rm ADM}=2M$ independently of $\alpha$, whereas the fixed-horizon HM4 configurations have $M_{\rm ADM}=M$. Table~\ref{tab:adm_masses} therefore provides the necessary conversion between the seed-normalized quantities of Table~\ref{tab:photon_rings_and_BH_Shadows} and the ADM-normalized quantities entering the angular comparison. 

\par The consequences of this normalization are displayed separately for the four geometries in Tables~\ref{tab:SGL_metric1}--\ref{tab:SGL_metric4}. The Schwarzschild reference gives descriptive central-value differences of approximately $6\%$ for M87* and $4\%$ for Sgr~A*, where these values serve as measures of angular-scale displacement. For HM1, Table~\ref{tab:SGL_metric1} shows that ADM normalization changes the interpretation suggested by the raw seed-normalized impact parameter: although $b_{\rm ph}$ grows in Table~\ref{tab:photon_rings_and_BH_Shadows}, the controlled points $\alpha=1$ and $\alpha=4$ have $\bar b_{\rm ph}=4.8813$ and $\bar b_{\rm ph}=4.6317$, respectively; these are both below the Schwarzschild value of $\bar b_{\rm ph}=5.1962$. The corresponding central-value differences increase to approximately $12\%$ and $16\%$ for M87*, with similar behavior for Sgr~A*. The exploratory $\alpha=10$ point is non-monotonic in the ADM-normalized comparison, returning to $\bar b_{\rm ph}=5.1149$ and correspondingly smaller percentage differences. Thus, the apparent monotonic growth of the seed-normalized shadow scale does not survive ADM normalization across the extended HM1 sequence. 

\par For HM2, Table~\ref{tab:SGL_metric2} displays a stronger ADM-normalization effect. The seed-normalized $b_{\rm ph}$ changes only mildly, whereas the impact parameter falls from $\bar b_{\rm ph}=4.8782$ at $\alpha=1$ to $\bar b_{\rm ph}=3.1702$ at $\alpha=10$. The corresponding M87* central-value difference rises from
approximately $12\%$ to $43\%$, with a comparable trend for Sgr~A*. The $\alpha=20$ configuration extends this behavior but lies outside the adopted controlled comparison range. This illustrates that a weak variation of the geometric impact parameter at fixed seed mass need not translate into a weak variation at fixed observed mass. For HM3, Table~\ref{tab:SGL_metric3} makes the importance of $M_{\rm ADM}=2M$ especially transparent. The comparatively large seed-normalized values of $b_{\rm ph}$ in Table~\ref{tab:photon_rings_and_BH_Shadows} become $\bar b_{\rm ph}=3.8999$, $3.7977$, and $3.6942$ for $\alpha=1,2,3$, respectively. The corresponding critical angular diameters lie well below the Schwarzschild reference, with descriptive central-value differences of approximately $28-33\%$ across the two sources. All three configurations lie within the DEC-admissible range, so this behavior characterizes the controlled HM3 fixed-horizon branch rather than an extrapolation beyond it. 

\par For HM4, Table~\ref{tab:SGL_metric4} shows a more modest reduction of the ADM-normalized impact parameter because the fixed-horizon condition gives $M_{\rm ADM}=M$. The values decrease from $\bar b_{\rm ph}=5.0934$ at $\alpha=1$ to $\bar b_{\rm ph}=4.9016$ at $\alpha=3$, accompanied by increasing central-value differences from the adopted EHT angular scales. Since all of these $\alpha>0$ HM4 configurations lie outside the DEC-admissible sector, Table~\ref{tab:SGL_metric4} should be interpreted solely as an illustration of how the HM4 geometric deformation would affect the angular scale.

\par Taken together, Tables~\ref{tab:adm_masses} and \ref{tab:SGL_metric1}--\ref{tab:SGL_metric4} demonstrate that ADM normalization is essential for any comparison performed at fixed observed mass. They also show that the response is not universal: HM1 becomes non-monotonic over the extended sequence, HM2 develops a substantial normalization-induced reduction in its angular scale, HM3 is strongly affected by its fixed $M_{\rm ADM}=2M$, and HM4 exhibits a comparatively mild trend. We reiterate that the percentages reported in these tables remain descriptive central-value differences; the geometrically predicted angular diameter and the observed emission ring diameter are not identical observables, and the tabulated values therefore do not constitute direct EHT constraints on $\alpha$ or $j$. 

\par The scalar QNM spectra provide a complementary probe of the same geometries. Table~\ref{tab:wkb_qnm_models} presents the sixth-order WKB frequencies for $\alpha=1$, $\ell=2,\ldots,7$, and $n=0,1,2$, with a common seed-mass normalization $M=1$. These were computed following Ref. \cite{Konoplya:2004ip}. Across this table, HM2 remains close to the Schwarzschild QNFs, whereas HM1 shows systematically smaller real frequencies and damping magnitudes. HM3 exhibits the largest departure among the controlled branches: for example, for the fundamental $\ell=2$ mode, the WKB result changes from $M\omega=0.4836-0.0968i$ for Schwarzschild to $M\omega=0.3213-0.0449i$ for HM3. HM4 instead gives moderately larger real frequencies and larger damping magnitudes than Schwarzschild for the exploratory $\alpha=1$ configuration. The RN reference lies above Schwarzschild in the real part over the modes shown and serves only as a familiar comparison geometry; we emphasize that its electromagnetic charge should not be identified with the effective charge appearing in the hairy solutions.

\par For the improved AIM calculation, we derived the generic coefficient functions in Eq.~\eqref{eq:generalAIM} following Refs. \cite{Cho:2009cj,Cho:2011sf}, after explicitly factoring the ingoing and outgoing QNM boundary behavior. The resulting formulation applies to the non-extremal, static, spherically symmetric and asymptotically flat geometries considered in this work, written in Schwarzschild-like coordinates, provided that the corresponding asymptotic factorization is well defined. The expansion point is chosen from the maximum of the effective potential, providing a uniform prescription across the four metrics. The improved AIM QNFs are recorded in  Table~\ref{tab:aim_qnm_models} for seed-mass $M=1$ and under the fixed-horizon $r_{\rm h}=2M$ constraint; these QNFs reproduce the same metric-dependent hierarchy. In particular, the strong reduction of both oscillation and damping scales for HM3, the comparatively small shift of HM2, the reduced frequencies of HM1, and the modest enhancement for HM4 persist throughout the listed multipoles and overtones. The AIM results therefore support the qualitative trends inferred from the WKB calculation without changing their physical interpretation as scalar test-field QNMs on fixed backgrounds. The level of agreement between the two calculations is quantified directly in Table~\ref{tab:WKB_AIM_PERCENT_ERROR}. Most entries agree to the precision displayed, and the largest listed discrepancy is approximately $0.33\%$, occurring in the imaginary part of the HM1 $(\ell,n)=(2,2)$ mode. It is important to note that this mode is among the less favorable cases for a WKB treatment because the overtone number is comparable to the angular multipole. The small discrepancies across the remaining modes provide a useful consistency check on the targeted roots. 

\par We do, however, acknowledge that since the WKB QNFs are used as seed values for the AIM root search, the two calculations are not completely independent at the level of root selection. The AIM QNFs are nevertheless obtained by solving the AIM quantization condition itself, with the WKB values serving only as starting estimates for the complex root finder. For this reason, the close agreement remains informative as a consistency check on the targeted QNM roots and on the metric-dependent trends across the modes considered. Table~\ref{tab:WKB_AIM_PERCENT_ERROR} should consequently be read as a cross-method consistency comparison rather than an absolute numerical-error estimate. A fully independent accuracy assessment would require an additional continued-fraction, spectral, or time-domain calculation; we reserve this for a future work.

\par The additional exploratory QNF calculations reported in Tables~\ref{tab:wkb_qnm_models_2}--\ref{tab:qnm_wkb_iaim_comparison} extend the comparison away from the common $r_{\rm h}=2M$ benchmark. Their purpose is distinct from that of Tables~\ref{tab:wkb_qnm_models}--\ref{tab:WKB_AIM_PERCENT_ERROR}: they test whether fixing the horizon radius alone is sufficient to determine the perturbative response of HM3 and HM4. These comparisons show that different choices of $(\alpha,j)$ can produce distinct scalar QNM spectra even when the horizon radius is held fixed. The QNM spectrum is therefore sensitive to the full radial form of the metric and cannot be characterized by $r_{\rm h}$ alone. For HM4, these extended configurations remain exploratory because relaxing the fixed-horizon condition does not restore DEC admissibility. This observation also clarifies the role of the fixed-horizon construction used throughout much of the analysis. Fixing $r_{\rm h}=2M$ provides a useful common geometric benchmark in seed-mass units and prevents changes in the coordinate horizon radius from obscuring the response to the deformation parameters. However, it is important to recognize that setting $r_{\rm h}=2M$ does not fix the ADM mass across the different families and should therefore not be interpreted as a comparison across objects of equal mass. This distinction is made explicit in Table~\ref{tab:adm_masses} and is essential for the angular comparisons in Tables~\ref{tab:SGL_metric1}--\ref{tab:SGL_metric4}. Such a distinction would likewise have to be taken into account if the seed-normalized QNFs in Tables~\ref{tab:wkb_qnm_models} and~\ref{tab:aim_qnm_models} were translated into phenomenological QNFs for black holes of fixed observed mass.

\par Overall, the tabulated results reveal a coherent but strongly metric-dependent picture. Table~\ref{tab:photon_rings_and_BH_Shadows} establishes how the deformations alter the null geodesic structure in seed-mass units; Table~\ref{tab:adm_masses} determines how those results must be renormalized for fixed-mass comparisons; Tables~\ref{tab:SGL_metric1}--\ref{tab:SGL_metric4} show the resulting changes in the characteristic angular scale; Tables~\ref{tab:wkb_qnm_models}, \ref{tab:aim_qnm_models}, and \ref{tab:WKB_AIM_PERCENT_ERROR} demonstrate the corresponding metric-dependent scalar perturbative response and the consistency of the two computational approaches. The auxiliary QNF comparisons of Appendix \ref{appendix} further emphasize that equal horizon radii do not imply equal spectra. No single diagnostic therefore provides a universal characterization of the deformation sector.

\par We reiterate that within the criteria adopted here, HM1--HM3 possess controlled parameter domains that can be analyzed consistently, whereas the $\alpha>0$ HM4 configurations remain exploratory because of their failure to satisfy the DEC. A complete assessment of the perturbative stability of the physically admissible solutions would require perturbing the metric together with the additional source sector. Likewise, a quantitative comparison with black hole imaging observations would require a consistent treatment of emission modeling and observational uncertainties. These extensions lie beyond the scope of the present work.

\section{Conclusion}
\label{Conclusion}

\par In this work, we investigated how the deformation parameters $\alpha$ and $j$ modify the geometric, optical, and perturbative properties of four gravitationally-decoupled hairy black hole metrics. Particular attention was given to identifying physically controlled parameter domains using the relevant energy conditions, horizon structure, and, where applicable, the RN-like interpretation of the parent DEC family. Within these criteria, HM1--HM3 possess controlled parameter ranges, whereas the nonzero-$\alpha$ HM4 configurations considered here are not DEC-admissible and are therefore retained only as exploratory geometric and perturbative benchmarks.

\par A common fixed-horizon construction, $r_{\rm h}=2M$, was used for much of the comparison in order to provide a uniform reference in seed-mass units and to separate changes in the exterior geometry from trivial changes in the coordinate horizon radius. We emphasize that this does not correspond to a comparison at fixed physical mass, because the ADM mass generally varies with the deformation parameters (as demonstrated explicitly in Table~\ref{tab:adm_masses}). This distinction becomes essential when translating the geometric results into angular observables corresponding to astrophysical black holes. In particular, the critical impact parameter must first be normalized by $M_{\rm ADM}$ before a fixed observed mass scale is introduced.

\par The photon sphere and shadow scale analysis shows that the response to the deformation is strongly metric dependent. HM1 exhibits an increase in both the photon sphere radius and the seed-normalized critical impact parameter over the configurations considered, while HM2 shows a more modest response; HM3 and HM4 exhibit decreasing trends along the corresponding benchmark branches. After ADM normalization, these trends need not be preserved directly. This is especially important for HM3, for which the fixed-horizon relation $j=2M/\alpha$ implies that $M_{\rm ADM}=2M$ throughout the branch. The resulting critical angular diameters therefore provide a useful illustration of the sensitivity of the shadow scale to the different deformations; we emphasize, however, that these should not be identified directly with the observed EHT emission ring diameters or interpreted as statistical constraints on the hair parameters. A quantitative observational test would additionally require the appropriate emission modeling, mass-distance uncertainties, and the systematic relation between the geometrical shadow scale and the measured image structure.

\par The scalar QNM spectra provide a complementary probe of the same background geometries. For the configurations considered, the effective potentials retain a smooth single barrier structure, with changes in their height and local curvature reflected in corresponding shifts of the oscillation frequencies and damping rates. The QNFs were calculated using the sixth-order WKB approximation and the improved AIM, with all frequencies reported in the common seed mass normalization $M=1$. Additional calculations performed beyond the fixed horizon condition show that configurations with the same horizon radius can nevertheless possess different scalar QNM spectra when the underlying values of $\alpha$ and $j$ differ. The horizon radius alone is therefore insufficient to characterize the perturbative response of these geometries. Across the multipoles and overtones considered, the WKB and improved AIM results are in close agreement. The WKB QNFs are used to initialize the AIM root search, so the two calculations are not fully independent at the level of root selection. On the other hand, the AIM QNFs are obtained from the AIM quantization condition itself, rather than on the initial estimate; as such, the resulting agreement provides a useful and dependable consistency check on both the targeted QNFs and the metric dependent trends identified in the spectra. We acknowledge that a fully independent assessment of the absolute numerical accuracy would require comparison with an additional method, such as a continued fraction, spectral, or time-domain analysis.

\par Taken together, our results show that the horizon structure, photon sphere properties, ADM-normalized shadow scales, effective potentials, and scalar test-field QNFs probe different aspects of the gravitationally-decoupled black hole geometries. Their combined behavior provides a more complete characterization of the deformation sector than any single diagnostic alone, while the physical interpretation remains sensitive to both the metric family and the admissibility of the corresponding parameter configuration. The scalar test-field analysis presented here provides information about perturbations of the fixed backgrounds, but a complete linear stability analysis would additionally require perturbing the geometry and the associated source sector. Similarly, a quantitative astrophysical interpretation of the optical and QNF results would require a dedicated phenomenological treatment at fixed observed mass, together with the relevant observational and modeling uncertainties. These questions provide natural directions for future work.

\acknowledgments
\noindent PAMG acknowledges the collaboration between De La Salle University - Manila and University of Johannesburg, which made this work possible. PAMG also thanks Department of Physics, De La Salle University-Manila, for its valuable mentorship and academic guidance, and the St. Miguel Febres Cordero Scholarship Program for its scholarship support. AC acknowledges the support of the Initiative Physique des Infinis (IPI), a research training programme of Idex SUPER at Sorbonne Université. ASC was partly supported by the National Research Foundation of South Africa, and thanks New York University - Abu Dhabi for its hospitality during the completion of this work. ER acknowledges the Department of Physics and the Office of the Vice President for Research and Innovation of De La Salle University for providing support that facilitated this research collaboration.

\appendix
\section{QNFs beyond the fixed-horizon branch}
\label{appendix}
\par In this Appendix, we present additional scalar QNF calculations that extend the analysis beyond the common fixed-horizon benchmark $r_{\rm h}=2M$ adopted in the main text. These configurations are included to examine the sensitivity of the QNM spectrum to the full set of deformation parameters when the horizon radius is allowed to vary from the Schwarzschild radius. With this in mind, we have set $r_{\rm h} = \alpha j = 3M$ for HM3 and $r_{\rm h} = 2M + \alpha j = 4M$ for HM4 (cf. Eq.~\eqref{eq:rhHM3} and Eq.~\eqref{eq:rhHM4}, respectively). In Tables~\ref{tab:wkb_qnm_models_2}, \ref{tab:aim_qnm_models_2}, and \ref{tab:WKB_AIM_PERCENT_ERROR_2}, we have set $\alpha=1$, such that $j_{\rm HM3}=3$ and $j_{\rm HM4}=2$. In Table~\ref{tab:qnm_wkb_iaim_comparison}, however, we have fixed $j=1$, such that $\alpha_{\rm HM3}=3$ and $\alpha_{\rm HM4}=2$.

\par Tables~\ref{tab:wkb_qnm_models_2} and~\ref{tab:aim_qnm_models_2} demonstrate the QNFs computed using the sixth-order WKB and AIM techniques, respectively, while Table~\ref{tab:WKB_AIM_PERCENT_ERROR_2} quantifies the small relative discrepancies between the two methods. The additional comparison in Table~\ref{tab:qnm_wkb_iaim_comparison} further illustrates that fixing the horizon radius alone does not uniquely determine the scalar QNM spectrum, since distinct choices of the deformation parameters can lead to different perturbative responses. 

\par There are two main comparisons to consider in this Appendix, the first of which is the performance of the WKB method and the AIM. Table~\ref{tab:WKB_AIM_PERCENT_ERROR_2} shows extremely good agreement for the $\alpha=1$ case. For HM3, there are no discrepencies evident at the working precision in the real part of the QNF; for the imaginary part, the largest relative error is ~0.05\%. For HM4, the largest discrepancies are 0.05\% in the real part and 0.12\% in the imaginary. The largest Table~\ref{tab:qnm_wkb_iaim_comparison} discrepancies occur when the multipoles are small and equivalent to or exceeded by the overtone number, as one would expect for a WKB calculation: for example, the (2,2) mode for HM3 reaches an error of 0.21\% in the imaginary part.  Most entries are nevertheless 0.00\% at the quoted precision, including essentially the entirety of the larger-$\ell$ sector. 

\par The second comparison concerns only HM3 and HM4, where $\alpha=1$ and $j\neq1$ for Tables~\ref{tab:wkb_qnm_models_2}--\ref{tab:WKB_AIM_PERCENT_ERROR_2}, and $\alpha\neq1$ and $j=1$ for Table~\ref{tab:qnm_wkb_iaim_comparison}. In both cases, the horizon radii are held fixed at $r_{\rm h}=3M$ for HM3 and $r_{\rm h}=4M$ for HM4. Thus, the comparison isolates the effect of changing the individual values of $\alpha$ and $j$ while preserving the corresponding horizon location. For HM3, this produces a small but systematic shift in the spectrum. Relative to the $\alpha=1$ results, the real part of the $\alpha\neq1$ QNF increases by approximately $2-3\%$, while the magnitude of the imaginary part increases by approximately $2\%$. The effect is somewhat larger for the higher overtones at low $\ell$: for example, the real part of the $(2,2)$ mode changes from $0.2314$ to $0.2390$, corresponding to a shift of approximately $3.3\%$.

\par HM4 is considerably less sensitive to the same change of parameter prescription. The real parts of the QNFs in Table~\ref{tab:qnm_wkb_iaim_comparison} differ from those obtained with $\alpha=1$ by only approximately $0.2\%$, while the imaginary parts change by approximately $0.5-0.6\%$. This behavior persists across the range of multipoles and overtones considered. The comparison therefore shows that, although fixing the horizon radius considerably restricts the resulting spectrum, it does not in general remove the dependence on the individual values of $\alpha$ and $j$. This dependence is more pronounced for HM3 than for HM4. Note also that the close agreement between the WKB and AIM calculations is maintained under both parameter choices, reinforcing the stability of the methods in the wake of changes to background parameters.

\begin{table}[htbp]
\centering
\caption{\textit{Sixth-order WKB QNFs $M\omega$ of a massless scalar test-field for the hairy black hole geometries, evaluated at $\alpha=1$ and seed mass $M=1$. For HM3, $r_{\rm h}=\alpha j=3M$, while for HM4, $r_{\rm h}=2M+\alpha j=4M$. These configurations therefore lie outside the common $r_{\rm h}=2M$ benchmark used in the main comparison.
}}
\label{tab:wkb_qnm_models_2}
\renewcommand{\arraystretch}{1.5}
\setlength{\tabcolsep}{10pt}

\resizebox{\textwidth}{!}{%
\begin{tabular}{|c|c|c|c|c|c|c|c|c|c|}
\hline
\multirow{2}{*}{\text{$\ell$}} & \multirow{2}{*}{\text{$n$}}
& \multicolumn{2}{c|}{\text{HM1}}
& \multicolumn{2}{c|}{\text{HM2}}
& \multicolumn{2}{c|}{\text{HM3}}
& \multicolumn{2}{c|}{\text{HM4}} \\
\cline{3-10}
& 
& \text{Re $\omega$} & \text{Im $\omega$}
& \text{Re $\omega$} & \text{Im $\omega$}
& \text{Re $\omega$} & \text{Im $\omega$}
& \text{Re $\omega$} & \text{Im $\omega$} \\
\hline

\multirow{3}{*}{2}
& 0   & 0.4539 & $-0.0824$
      & 0.4820 & $-0.0966$
      & 0.2485 & $-0.0369$
      & 0.2424 & $-0.0486$ \\
& 1   & 0.4421 & $-0.2504$
      & 0.4609 & $-0.2950$
      & 0.2426 & $-0.1116$
      & 0.2324 & $-0.1487$ \\
& 2   & 0.4249 & $-0.4230$
      & 0.4236 & $-0.5078$
      & 0.2314 & $-0.1891$
      & 0.2156 & $-0.2560$ \\
\hline

\multirow{3}{*}{3}
& 0   & 0.6336 & $-0.0822$
      & 0.6734 & $-0.0964$
      & 0.3472 & $-0.0369$
      & 0.3385 & $-0.0485$ \\
& 1   & 0.6243 & $-0.2482$
      & 0.6579 & $-0.2919$
      & 0.3429 & $-0.1110$
      & 0.3311 & $-0.1470$ \\
& 2   & 0.6079 & $-0.4177$
      & 0.6286 & $-0.4954$
      & 0.3347 & $-0.1866$
      & 0.3174 & $-0.2494$ \\
\hline

\multirow{3}{*}{4}
& 0   & 0.8136 & $-0.0821$
      & 0.8651 & $-0.0963$
      & 0.4460 & $-0.0368$
      & 0.4348 & $-0.0485$ \\
& 1   & 0.8061 & $-0.2473$
      & 0.8528 & $-0.2907$
      & 0.4427 & $-0.1108$
      & 0.4289 & $-0.1462$ \\
& 2   & 0.7921 & $-0.4149$
      & 0.8292 & $-0.4899$
      & 0.4361 & $-0.1856$
      & 0.4177 & $-0.2465$ \\
\hline

\multirow{3}{*}{5}
& 0   & 0.9937 & $-0.0821$
      & 1.0569 & $-0.0963$
      & 0.5449 & $-0.0368$
      & 0.5311 & $-0.0484$ \\
& 1   & 0.9875 & $-0.2469$
      & 1.0468 & $-0.2899$
      & 0.5422 & $-0.1106$
      & 0.5263 & $-0.1459$ \\
& 2   & 0.9756 & $-0.4134$
      & 1.0271 & $-0.4870$
      & 0.5368 & $-0.1850$
      & 0.5169 & $-0.2450$ \\
\hline

\multirow{3}{*}{6}
& 0   & 1.1739 & $-0.0821$
      & 1.2488 & $-0.0963$
      & 0.6438 & $-0.0368$
      & 0.6275 & $-0.0484$ \\
& 1   & 1.1686 & $-0.2467$
      & 1.2402 & $-0.2896$
      & 0.6415 & $-0.1106$
      & 0.6234 & $-0.1457$ \\
& 2   & 1.1583 & $-0.4125$
      & 1.2233 & $-0.4854$
      & 0.6369 & $-0.1848$
      & 0.6153 & $-0.2441$ \\
\hline

\multirow{3}{*}{7}
& 0   & 1.3543 & $-0.0821$
      & 1.4407 & $-0.0963$
      & 0.7427 & $-0.0368$
      & 0.7239 & $-0.0484$ \\
& 1   & 1.3496 & $-0.2465$
      & 1.4332 & $-0.2894$
      & 0.7407 & $-0.1105$
      & 0.7203 & $-0.1455$ \\
& 2   & 1.3405 & $-0.4119$
      & 1.4185 & $-0.4843$
      & 0.7367 & $-0.1846$
      & 0.7133 & $-0.2436$ \\
\hline
\end{tabular}}
\end{table}

\begin{table}[t]
\centering
\caption{\textit{Improved AIM scalar QNFs $M\omega$ for the hairy black hole geometries, evaluated at $\alpha=1$ and seed mass $M=1$.  For HM3, $r_{\rm h}=\alpha j=3M$, while for HM4, $r_{\rm h}=2M+\alpha j=4M$. These configurations therefore lie outside the common $r_{\rm h}=2M$ benchmark used in the main comparison.}}
\label{tab:aim_qnm_models_2}
\vskip 0.2cm
\renewcommand{\arraystretch}{1.5}
\setlength{\tabcolsep}{10pt}

\resizebox{\textwidth}{!}{%
\begin{tabular}{|c|c|c|c|c|c|c|c|c|c|}
\hline
\multirow{2}{*}{\text{$\ell$}} & \multirow{2}{*}{\text{$n$}}
& \multicolumn{2}{c|}{\text{HM1}}
& \multicolumn{2}{c|}{\text{HM2}}
& \multicolumn{2}{c|}{\text{HM3}}
& \multicolumn{2}{c|}{\text{HM4}} \\
\cline{3-10}
& 
& \text{Re $\omega$} & \text{Im $\omega$}
& \text{Re $\omega$} & \text{Im $\omega$}
& \text{Re $\omega$} & \text{Im $\omega$}
& \text{Re $\omega$} & \text{Im $\omega$} \\
\hline

\multirow{3}{*}{2}
& 0   & 0.4539 & $-0.0824$
      & 0.4821 & $-0.0966$
      & 0.2485 & $-0.0369$
      & 0.2424 & $-0.0486$ \\
& 1   & 0.4419 & $-0.2502$
      & 0.4609 & $-0.2950$
      & 0.2426 & $-0.1116$
      & 0.2324 & $-0.1486$ \\
& 2   & 0.4242 & $-0.4244$
      & 0.4237 & $-0.5076$
      & 0.2314 & $-0.1890$
      & 0.2157 & $-0.2557$ \\
\hline

\multirow{3}{*}{3}
& 0   & 0.6336 & $-0.0822$
      & 0.6734 & $-0.0964$
      & 0.3472 & $-0.0369$
      & 0.3385 & $-0.0485$ \\
& 1   & 0.6242 & $-0.2481$
      & 0.6579 & $-0.2919$
      & 0.3429 & $-0.1110$
      & 0.3311 & $-0.1469$ \\
& 2   & 0.6079 & $-0.4179$
      & 0.6286 & $-0.4954$
      & 0.3347 & $-0.1866$
      & 0.3174 & $-0.2494$ \\
\hline

\multirow{3}{*}{4}
& 0   & 0.8136 & $-0.0821$
      & 0.8651 & $-0.0963$
      & 0.4460 & $-0.0368$
      & 0.4348 & $-0.0485$ \\
& 1   & 0.8061 & $-0.2473$
      & 0.8528 & $-0.2907$
      & 0.4427 & $-0.1108$
      & 0.4289 & $-0.1462$ \\
& 2   & 0.7921 & $-0.4150$
      & 0.8292 & $-0.4899$
      & 0.4361 & $-0.1856$
      & 0.4177 & $-0.2465$ \\
\hline

\multirow{3}{*}{5}
& 0   & 0.9937 & $-0.0821$
      & 1.0569 & $-0.0962$
      & 0.5449 & $-0.0368$
      & 0.5311 & $-0.0484$ \\
& 1   & 0.9875 & $-0.2469$
      & 1.0468 & $-0.2899$
      & 0.5422 & $-0.1106$
      & 0.5263 & $-0.1459$ \\
& 2   & 0.9756 & $-0.4135$
      & 1.0271 & $-0.4870$
      & 0.5368 & $-0.1850$
      & 0.5169 & $-0.2450$ \\
\hline

\multirow{3}{*}{6}
& 0   & 1.1739 & $-0.0821$
      & 1.2488 & $-0.0963$
      & 0.6438 & $-0.0368$
      & 0.6275 & $-0.0484$ \\
& 1   & 1.1686 & $-0.2467$
      & 1.2402 & $-0.2896$
      & 0.6415 & $-0.1106$
      & 0.6234 & $-0.1457$ \\
& 2   & 1.1583 & $-0.4126$
      & 1.2233 & $-0.4854$
      & 0.6369 & $-0.1847$
      & 0.6153 & $-0.2441$ \\
\hline

\multirow{3}{*}{7}
& 0   & 1.3543 & $-0.0821$
      & 1.4407 & $-0.0963$
      & 0.7427 & $-0.0368$
      & 0.7239 & $-0.0484$ \\
& 1   & 1.3496 & $-0.2465$
      & 1.4332 & $-0.2894$
      & 0.7407 & $-0.1105$
      & 0.7203 & $-0.1455$ \\
& 2   & 1.3405 & $-0.4119$
      & 1.4185 & $-0.4843$
      & 0.7367 & $-0.1846$
      & 0.7133 & $-0.2436$ \\ 
\hline
\end{tabular}}
\end{table}

\begin{table}[t]
\centering
\caption{\textit{Relative percentage discrepancies between the WKB and improved AIM scalar QNFs reported in Tables~\ref{tab:wkb_qnm_models_2} and~\ref{tab:aim_qnm_models_2}, respectively. The real and imaginary parts are evaluated separately using the WKB result as the reference denominator. Entries reported as $0.00\%$ indicate agreement at the numerical precision used to construct the table and should not be interpreted as exact equality.
}}
\label{tab:WKB_AIM_PERCENT_ERROR_2}
\vskip 0.2cm
\renewcommand{\arraystretch}{1.5}
\setlength{\tabcolsep}{7.5pt}

\begin{tabular}{|c|c|c|c|c|c|c|c|c|c|}
\hline
\multirow{2}{*}{\text{$\ell$}} & \multirow{2}{*}{\text{$n$}}
& \multicolumn{2}{c|}{\text{HM1}}
& \multicolumn{2}{c|}{\text{HM2}}
& \multicolumn{2}{c|}{\text{HM3}}
& \multicolumn{2}{c|}{\text{HM4}} \\
\cline{3-10}
&
& \text{Re $\omega$} & \text{Im $\omega$}
& \text{Re $\omega$} & \text{Im $\omega$}
& \text{Re $\omega$} & \text{Im $\omega$}
& \text{Re $\omega$} & \text{Im $\omega$} \\
\hline

\multirow{3}{*}{2}
& 0 & 0.00 & $0.00$
      & 0.02 & $0.00$
      & 0.00 & $0.00$
      & 0.00 & $0.00$ \\
& 1 & 0.05 & $0.08$
      & 0.00 & $0.00$
      & 0.00 & $0.00$
      & 0.00 & $0.07$ \\
& 2 & 0.16 & $0.33$
      & 0.02 & $0.04$
      & 0.00 & $0.05$
      & 0.05 & $0.12$ \\
\hline

\multirow{3}{*}{3}
& 0 & 0.00 & $0.00$
      & 0.00 & $0.00$
      & 0.00 & $0.00$
      & 0.00 & $0.00$ \\
& 1 & 0.02 & $0.04$
      & 0.00 & $0.00$
      & 0.00 & $0.00$
      & 0.00 & $0.07$ \\
& 2 & 0.00 & $0.05$
      & 0.00 & $0.00$
      & 0.00 & $0.00$
      & 0.00 & $0.00$ \\
\hline

\multirow{3}{*}{4}
& 0 & 0.00 & $0.00$
      & 0.00 & $0.00$
      & 0.00 & $0.00$
      & 0.00 & $0.00$ \\
& 1 & 0.00 & $0.00$
      & 0.00 & $0.00$
      & 0.00 & $0.00$
      & 0.00 & $0.00$ \\
& 2 &  0.00 & $0.02$
      & 0.00 & $0.00$
      & 0.00 & $0.00$
      & 0.00 & $0.00$ \\
\hline

\multirow{3}{*}{5}
& 0 & 0.00 & $0.00$
      & 0.00 & $0.10$
      & 0.00 & $0.00$
      & 0.00 & $0.00$ \\
& 1 & 0.00 & $0.00$
      & 0.00 & $0.00$
      & 0.00 & $0.00$
      & 0.00 & $0.00$ \\
& 2 & 0.00 & $0.02$
      & 0.00 & $0.00$
      & 0.00 & $0.00$
      & 0.00 & $0.00$ \\
\hline

\multirow{3}{*}{6}
& 0 & 0.00 & $0.00$
      & 0.00 & $0.00$
      & 0.00 & $0.00$
      & 0.00 & $0.00$ \\
& 1 & 0.00 & $0.00$
      & 0.00 & $0.00$
      & 0.00 & $0.00$
      & 0.00 & $0.00$ \\
& 2 & 0.00 & $0.02$
      & 0.00 & $0.00$
      & 0.00 & $0.05$
      & 0.00 & $0.00$ \\
\hline

\multirow{3}{*}{7}
& 0 & 0.00 & $0.00$
      & 0.00 & $0.00$
      & 0.00 & $0.00$
      & 0.00 & $0.00$ \\
& 1 & 0.00 & $0.00$
      & 0.00 & $0.00$
      & 0.00 & $0.00$
      & 0.00 & $0.00$ \\
& 2 & 0.00 & $0.00$
      & 0.00 & $0.00$
      & 0.00 & $0.00$
      & 0.00 & $0.00$ \\
\hline
\end{tabular}
\end{table}

\begin{table}[t]
\centering
\caption{\textit{Comparison of the QNFs obtained from the WKB and improved AIM methods for HM3 and HM4, together with the percentage error. Here, we fix $M=1$ and $j = 1$, and impose that $r_{\rm h} = \alpha j =3M$ for HM3 and $r_{\rm h} = 2M + \alpha j =4M$ for HM4. Relative percentage error entries reported as $0.00\%$ indicate agreement at the numerical precision used to construct the table and should not be interpreted as exact equality.}}
\label{tab:qnm_wkb_iaim_comparison}
\vskip 0.2cm
\renewcommand{\arraystretch}{1.8}
\setlength{\tabcolsep}{5pt}

\resizebox{\linewidth}{!}{%
\begin{tabular}{|c|c|c|c|c|c|c|c|c|c|c|c|c|c|}
\hline
\multirow{3}{*}{\text{$L$}} & \multirow{3}{*}{\text{$n$}}
& \multicolumn{2}{c|}{\multirow{2}{*}{\text{HM3 (WKB)}}}
& \multicolumn{2}{c|}{\multirow{2}{*}{\text{HM4 (WKB)}}}
& \multicolumn{2}{c|}{\multirow{2}{*}{\text{HM3 (AIM)}}}
& \multicolumn{2}{c|}{\multirow{2}{*}{\text{HM4 (AIM)}}}
& \multicolumn{4}{c|}{\text{Rel. percentage error}} \\
\cline{11-14}
& & \multicolumn{2}{c|}{} 
& \multicolumn{2}{c|}{}
& \multicolumn{2}{c|}{}
& \multicolumn{2}{c|}{}
& \multicolumn{2}{c|}{\text{HM3}}
& \multicolumn{2}{c|}{\text{HM4}} \\
\cline{3-14}
& &
\text{Re $\omega$} & \text{Im $\omega$}
& \text{Re $\omega$} & \text{Im $\omega$}
& \text{Re $\omega$} & \text{Im $\omega$}
& \text{Re $\omega$} & \text{Im $\omega$}
& \text{Re $\omega$} & \text{Im $\omega$}
& \text{Re $\omega$} & \text{Im $\omega$} \\
\hline

\multirow{3}{*}{2}
& 0 & 0.2531 & $-0.0377$ 
    & 0.2430 & $-0.0489$ 
    & 0.2531 & $-0.0377$ 
    & 0.2430 & $-0.0489$ 
    & 0.00 & $0.00$
    & 0.00 & $0.00$ \\
& 1 & 0.2481 & $-0.1140$ 
    & 0.2328 & $-0.1495$ 
    & 0.2481 & $-0.1139$ 
    & 0.2329 & $-0.1495$ 
    & 0.00 & $0.09$
    & 0.04 & $0.00$ \\
& 2 & 0.2390 & $-0.1932$ 
    & 0.2160 & $-0.2576$ 
    & 0.2390 & $-0.1928$ 
    & 0.2160 & $-0.2572$ 
    & 0.04 & $0.21$
    & 0.09 & $0.16$ \\
\hline

\multirow{3}{*}{3}
& 0 & 0.3535 & $-0.0376$ 
    & 0.3393 & $-0.0488$ 
    & 0.3535 & $-0.0376$ 
    & 0.3393 & $-0.0488$ 
    & 0.00 & $0.00$
    & 0.00 & $0.00$ \\
& 1 & 0.3499 & $-0.1133$ 
    & 0.3318 & $-0.1478$ 
    & 0.3499 & $-0.1133$ 
    & 0.3318 & $-0.1478$ 
    & 0.00 & $0.00$
    & 0.00 & $0.00$ \\
& 2 & 0.3430 & $-0.1905$ 
    & 0.3180 & $-0.2508$ 
    & 0.3430 & $-0.1904$ 
    & 0.3180 & $-0.2508$ 
    & 0.00 & $0.05$
    & 0.00 & $0.00$ \\
\hline

\multirow{3}{*}{4}
& 0 & 0.4541 & $-0.0376$ 
    & 0.4358 & $-0.0488$ 
    & 0.4541 & $-0.0376$ 
    & 0.4358 & $-0.0487$ 
    & 0.00 & $0.00$
    & 0.00 & $0.20$ \\
& 1 & 0.4513 & $-0.1130$ 
    & 0.4299 & $-0.1470$ 
    & 0.4513 & $-0.1130$ 
    & 0.4299 & $-0.1470$ 
    & 0.00 & $0.00$
    & 0.00 & $0.00$ \\
& 2 & 0.4457 & $-0.1894$ 
    & 0.4186 & $-0.2479$ 
    & 0.4457 & $-0.1894$ 
    & 0.4186 & $-0.2479$ 
    & 0.00 & $0.00$
    & 0.00 & $0.00$ \\
\hline

\multirow{3}{*}{5}
& 0 & 0.5548 & $-0.0376$ 
    & 0.5324 & $-0.0487$ 
    & 0.5548 & $-0.0376$ 
    & 0.5324 & $-0.0487$ 
    & 0.00 & $0.00$
    & 0.00 & $0.00$ \\
& 1 & 0.5524 & $-0.1129$ 
    & 0.5275 & $-0.1467$ 
    & 0.5524 & $-0.1129$ 
    & 0.5275 & $-0.1467$ 
    & 0.00 & $0.00$
    & 0.00 & $0.00$ \\
& 2 & 0.5478 & $-0.1888$ 
    & 0.5180 & $-0.2464$ 
    & 0.5478 & $-0.1888$ 
    & 0.5180 & $-0.2463$ 
    & 0.00 & $0.00$
    & 0.00 & $0.04$ \\
\hline

\multirow{3}{*}{6}
& 0 & 0.6555 & $-0.0376$ 
    & 0.6290 & $-0.0487$ 
    & 0.6555 & $-0.0376$ 
    & 0.6290 & $-0.0487$ 
    & 0.00 & $0.00$
    & 0.00 & $0.00$ \\
& 1 & 0.6535 & $-0.1128$ 
    & 0.6248 & $-0.1465$ 
    & 0.6535 & $-0.1128$ 
    & 0.6248 & $-0.1465$ 
    & 0.00 & $0.00$
    & 0.00 & $0.00$ \\
& 2 & 0.6495 & $-0.1885$ 
    & 0.6167 & $-0.2455$ 
    & 0.6495 & $-0.1885$ 
    & 0.6167 & $-0.2455$ 
    & 0.00 & $0.00$
    & 0.00 & $0.00$ \\
\hline

\multirow{3}{*}{7}
& 0 & 0.7562 & $-0.0376$ 
    & 0.7256 & $-0.0487$ 
    & 0.7562 & $-0.0376$ 
    & 0.7256 & $-0.0487$ 
    & 0.00 & $0.00$
    & 0.00 & $0.06$ \\
& 1 & 0.7545 & $-0.1128$ 
    & 0.7220 & $-0.1463$ 
    & 0.7545 & $-0.1128$ 
    & 0.7220 & $-0.1463$ 
    & 0.01 & $0.00$
    & 0.00 & $0.00$ \\
& 2 & 0.7510 & $-0.1883$ 
    & 0.7149 & $-0.2449$ 
    & 0.7510 & $-0.1883$ 
    & 0.7149 & $-0.2449$ 
    & 0.00 & $0.00$
    & 0.00 & $0.00$ \\
\hline
\end{tabular}}
\end{table}

\newpage
\bibliographystyle{aipnum4-2} 
\bibliography{article_hBHs}

\end{document}